\documentclass[stslayout, noinfoline]{imsart}
\usepackage{geometry} 
\usepackage{graphicx}	
\usepackage{amssymb, amsmath}
\usepackage{natbib}
\usepackage[usenames,dvipsnames]{xcolor}
\usepackage[colorlinks, linkcolor=MidnightBlue, citecolor=MidnightBlue, urlcolor=MidnightBlue]{hyperref}

\usepackage[linesnumbered]{algorithm2e}
\usepackage{algpseudocode}
\RestyleAlgo{ruled}

\usepackage{subfigure}
\usepackage{verbatimbox}
\usepackage{tikz}
\usetikzlibrary{calc, arrows.meta}
\usepackage{pgfmath}

\usepackage{systeme}

\usepackage{amsmath}
\usepackage{amsthm}

\theoremstyle{definition}

\newtheorem{remark}{Remark}
\newtheorem{lemma}{Lemma}
\newtheorem{proposition}{Proposition}

\begin{document}

\begin{frontmatter}

\title{General adjoint-differentiated Laplace approximation}
\runtitle{Adjoint-differentiated Laplace}

\begin{aug}
\author{Charles C. Margossian$^\dagger$}
  \runauthor{Margossian}
  \address{$^\dagger$Flatiron Institute, Center for Computational Mathematics.
  This article originally appeared in the author's PhD thesis \citep{Margossian:2022}. 
  Work mostly done  while at Columbia University, Department of Statistics. This version contains some minor revisions.
  Contact: cmargossian@flatironinstitute.org}
\end{aug}

\begin{abstract}
The hierarchical prior used in Latent Gaussian models (LGMs) induces a posterior geometry prone to frustrate inference algorithms.
Marginalizing out the latent Gaussian variable using an integrated Laplace approximation removes the offending geometry, allowing us to do efficient inference on the hyperparameters.
To use gradient-based inference we need to compute the approximate marginal likelihood and its gradient.
The adjoint-differentiated Laplace approximation differentiates the marginal likelihood and scales well with the dimension of the hyperparameters. 
While this method can be applied to LGMs with any prior covariance, it only works for likelihoods with a diagonal Hessian.
Furthermore, the algorithm requires methods which compute the first three derivatives of the likelihood with current implementations relying on analytical derivatives.
I propose a generalization which is applicable to a broader class of likelihoods and does not require analytical derivatives of the likelihood.
Numerical experiments suggest the added flexibility comes at no computational cost: on a standard LGM, the new method is in fact slightly faster than the existing adjoint-differentiated Laplace approximation.
I also apply the general method to an LGM with an unconventional likelihood.
This example highlights the algorithm's potential, as well as persistent challenges. \\ \ \\
\end{abstract}

\end{frontmatter}

\section{Introduction}

Latent Gaussian models (LGMs) are a popular class of Bayesian models,
which include Gaussian processes, multilevel models
and population models.
Their general formulation is
$$
    \phi  \sim  \pi(\phi), \qquad 
    \theta \sim \text{Normal}(0, K(\phi)), \qquad
    y  \sim  \pi(y \mid \theta, \eta),
$$
where $\phi \in \mathbb R^p$ and $\eta \in \mathbb R^T$ denote the hyperparameters
and $\theta \in \mathbb R^n$ is the latent Gaussian variable. 
Our goal is to compute the posterior distribution \mbox{$\pi(\phi, \eta, \theta \mid y)$}.

The hierarchical prior on $\theta$ induces a challenging posterior geometry,
specifically an uneven curvature produced by the interaction between $\phi$ and $\theta$, and a strong posterior correlation between the different components of $\theta$.
This geometry can frustrate inference algorithms, including Markov chain Monte Carlo \citep{Betancourt:2015} and variational inference \citep{Zhang:2021}.
The \textit{integrated Laplace approximation} marginalizes out $\theta$ using an approximation \citep{Tierney:1986}, thereby allowing us to do inference on the more manageable distribution $\pi(\phi, \eta \mid y)$ \citep{Rue:2009}.
The posterior distribution of $\theta$ can then be studied by approximating the conditional distribution $\pi(\theta \mid \phi, \eta, y)$ in a subsequent step.

Most existing implementations of the integrated Laplace approximation algorithmically restrict the class of LGMs we can fit.
This is because developers focus on specific motivating problems and
write algorithms whose speed and stability depend on certain regularity conditions.
To expand the scope of the integrated Laplace approximation, we must develop methods which do not rely on such conditions.
Furthermore the advent of automatic differentiation in Machine Learning and Computational Statistics presents new opportunities to write code which is both more general and more efficient \citep{Baydin:2018, Margossian:2019, Margossian:2022-autodiff}.

Beyond algorithmic limitations, there is also an inferential limitation, which is that the Laplace approximation may not be adequate.
It is worth noting that the quality of the approximation depends not only on the likelihood distribution but also on the interaction of the likelihood and the prior.
In the limiting case where there is no observation tied to a particular parameter $\theta_i$ (e.g. empty cell in spatial model),  the posterior distribution of $\theta_i$ is exactly normal.
If the data is sparse, it will be approximatively normal.
For examples on how the data regime influences the quality of the approximation,
see the discussions by \citet{Vanhatalo:2009} and \citet{Talts:2018}.

Here I merely address the problem of constructing and differentiating the Laplace approximation,
while recognizing that the approximation is not always useful.
This project has two immediate benefits:
\begin{itemize}
  \item[] (i) When writing software to support a menu of likelihoods, a single function can be used for all likelihoods, thereby making the code shorter, more readable and straightforward to expand.
  \item[]
  \item[] (ii) It is possible to experiment with new likelihoods and conduct research on the utility of the integrated Laplace approximation.
\end{itemize}
I achieve both of these goals by building a prototype of the method in the probabilistic programming language \texttt{Stan} \citep{Carpenter:2015, Carpenter:2017}.
A longer term goal is to support software where users specify their own likelihood, while providing diagnostics to flag cases where the approximation is not appropriate.

\subsection{Existing implementation and limitation}  \label{sec:classical-limitation}

Fast algorithms take advantage of the convenient structure found in classical models at the expanse of applications to less conventional cases.
For example, the seminal algorithms by \citet{Rasmussen:2006} for inference on Gaussian processes assume the following:
\begin{itemize}
  \item[] (i) $\eta = \emptyset$
  \item[]
  \item[] (ii) The log likelihood $\log \pi(y \mid \theta, \eta) = \log \pi(y \mid \theta)$ has a diagonal Hessian. This often means each observation $y_i$ can only depend on a single component of $\theta$.
  \item[]
  \item[] (iii) $\pi(y \mid \theta, \eta)$ is a log-concave likelihood and the Hessian is negative definite.
\end{itemize}
We can take advantage of these conditions to build a numerically stable Newton solver to find the mode of $\pi(\theta \mid y, \phi, \eta)$ when constructing the Laplace approximation.

Calculations of the approximate log marginal likelihood, $\log \pi_\mathcal{G}(y \mid \phi, \eta)$, and its gradient with respect to $\phi$ require methods to explicitly compute the derivative of the prior covariance, $\partial K / \partial \phi$,
and the first three derivatives of the likelihood with respect to $\theta$.
This means users must either pick from a menu of prior covariances and likelihoods,
with pre-coded derivatives, or engage in the time-consuming and error prone task of hand-coding the requisite derivatives.
Many software implementations inherit at least some of these limitations \citep[e.g][]{Vanhatalo:2013, Rue:2017, Margossian:2020-laplace}.

The following examples violate the above conditions and required adjustments to make the computation of the Laplace approximation feasible:

\noindent
\textit{Gaussian process regression with a Student-$t$ likelihood.} 
\citet{Vanhatalo:2009} and later \citet{Jylanki:2011} propose to use a Student-$t$ likelihood in order to make Gaussian process regression robust to outliers.
The Student-$t$ likelihood is parameterized by the latent Gaussian variable and an additional scale parameter, meaning $\eta \neq \emptyset$.
Furthermore the Student-$t$ distribution is not log-concave and thus its Hessian not  negative definite.

\noindent
\textit{Motorcyle Gaussian process example.}
This example, described by \citet{Tolvanen:2014} and \citet{Vehtari:2021}, combines two Gaussian processes,
\begin{eqnarray*}
  y & \sim & \text{normal}(\mu(x), \Sigma) \\
  \mu & \sim & \text{GP}(0, K_1(x)) \\
  \tau & \sim & \text{GP}(0, K_2(x)),
\end{eqnarray*}
where
\begin{equation*}
  \Sigma = \begin{pmatrix}
  \exp(\tau_1(x)) & 0 & \cdots  & \cdots & 0 \\
  0 & \exp(\tau_2(x)) & 0 & \cdots & 0 \\
  \vdots & \vdots & \ddots & \ddots & \vdots \\
  0 & \cdots & \cdots & \cdots & \exp(\tau_n(x))
  \end{pmatrix}
\end{equation*}
Both the mean and variance vary as a function of the covariate $x$.
Conditional on $\mu_i$ and $\tau_i$, the $y_i$'s are independent of the other elements of $y$.
Combining the latent Gaussian variables into one,
\begin{equation*}
  \theta = (\mu_1, \tau_1, \mu_2, \tau_2, \cdots, \mu_n, \tau_n),
\end{equation*}
the above reduces to a single Gaussian process.
The resulting model admits a $2 \times 2$ block-diagonal Hessian
and is typically not negative-definite.
The GPStuff package implements an integrated Laplace approximation for this example, using an optimizer with carefully tuned initializations \citep{Vanhatalo:2013}.

Additional examples of LGMs with a less conventional structure include population pharmacometrics models with a differential equation based likelihood \citep[e.g][]{Gibaldi:1982, Gastonguay:2013, Margossian:2021-torsten}, neural networks with a horseshoe prior and more.
To my knowledge, the integrated Laplace approximation has not been used for these or related examples.
Section~\ref{sec:gam-experiment} presents an attempt to use Hamiltonian Monte Carlo with an integrated Laplace approximation on a standard pharmacometrics model,
using the general implementation developed in this article.

\subsection{Existing methods}

Expanding the scope of the integrated Laplace approximation requires finding more general methods to (i) construct a Laplace approximation of  $\pi(\theta \mid \phi, \eta, y)$
which is fundamentally an optimization problem,
and (ii) differentiate the approximate marginal distribution 
\mbox{$\pi_\mathcal{G}(y \mid \phi, \eta)$}.

\subsubsection{Alternative Newton solvers.}

The Newton solver employed by \citet{Rasmussen:2006} assumes the negative Hessian
\begin{equation*}
  W \triangleq - \frac{\partial^2 \pi(y \mid \theta, \eta)}{\partial \theta^2}
\end{equation*}
is diagonal and positive definite, meaning $W_{ii} \ge 0$.
If the likelihood is not log concave this condition is violated.
To address this \citet{Vanhatalo:2009} and \citet{Rasmussen:2010} propose modified Newton solvers.
Another direction may be to use the Fisher information matrix rather than the Hessian.\footnote{Personal communication with Jarno Vanhatalo.}
The information matrix is always semi-positive definite 
while the definiteness of the Hessian depends on the value of $\theta$,
and crucially values of $\theta$ we encounter along the optimization path.

\subsubsection{Gradient computation.}

State-of-the-art implementations of the integrated Laplace approximation,
for example in \texttt{GPStuff} \citep{Vanhatalo:2013} and in \texttt{INLA} \citep{Rue:2017},
are (in part) based on the algorithms by \citet{Rasmussen:2006}.
These packages require methods which explicitly compute the derivatives, $\partial K / \partial \phi$,
and higher-order derivatives of $\log \pi(y \mid \theta, \eta)$ with respect to $\theta$ and $\eta$.
These derivatives are then plugged into a differentiation algorithm.
The optimization step which produces the Laplace approximation
and the differentiation of the approximate marginal distribution are done jointly,
meaning a host of terms can be carried over from one calculation to the other.
Most notably, the expansive Cholesky decomposition done during the final step of the Newton optimizer can be reused in the differentiation step.

\citet{Kristensen:2016} apply automatic differentiation to the Laplace approximation and remove the requirement for any analytical derivatives, implementing their flexible strategy in the \texttt{TMB} package.
Their \textit{inverse subset algorithm} bypasses the explicit computation of intermediate Jacobians---in this sense, it is as an adjoint method.

\citet{Margossian:2020-laplace} use the same principles of automatic differentiation to construct the \textit{adjoint-differentiated Laplace approximation} and show that the calculation of $\partial K / \partial \phi$ is both superfluous and prohibitively expansive for high-dimensional $\phi$.
We will see that a similar reasoning applies to derivatives of the likelihood.
%
%
The adjoint-differentiated Laplace approximation differs from the inverse subset algorithm in two ways.
First it requires analytical higher-order derivatives of the likelihood;
the main object of this article is to eliminate this requirement.
Secondly, the inverse subset algorithm treats the optimizer as a black box, while the adjoint-differentiated Laplace approximation jointly optimizes and differentiates, in line with the procedure by \citet{Rasmussen:2006}.

\subsection{Aim and results}

I present a generalization of the adjoint-differentiated Laplace approximation
for a class of Newton solvers, which have appeared in the literature \citep[e.g.][]{Rasmussen:2006, Vanhatalo:2009, Rasmussen:2010, Vanhatalo:2021}.
This unifying framework follows from the realization that most of these solvers are distinguished only by their application of the Woodburry-Sherman-Morrison matrix inversion lemma.
The key step to insure a numerically stable algorithm is to carefully choose a $B$-matrix which can  be safely inverted.
Judicious choices of $B$ depend on the properties of the Hessian and the prior covariance.
The cost of the Newton step is then dominated by an $\mathcal O(n^3)$ decomposition of $B$.

Automatic differentiation removes the requirement for analytical derivatives of the likelihood, provided all operations in the evaluation of the likelihood support both forward and reverse mode differentiation.
Differentiating the Laplace approximation requires propagating derivatives through the mode of $\log \pi(\theta \mid y, \phi, \eta)$.
This can be done using the implicit function theorem.
\citet{Gaebler:2021} shows that the cost of differentiation is then dominated by an LU decomposition.
This costly operation is eliminated by reusing the $B$-matrix, decomposed in the final Newton step.
As in \citet{Margossian:2020-laplace}, I use the adjoint method of automatic differentiation, taking care to adapt the method to higher-order derivatives.
Another important step for an efficient implementation is to exploit the sparsity of the Hessian.
Without loss of generality, suppose the Hessian, $\nabla^2_\theta \log \pi(y \mid \theta, \eta)$, is block diagonal, with block size $m \times m$.
The number of automatic differentiation sweeps required to differentiate the approximate marginal density, $\pi_\mathcal{G}(y \mid \theta, \eta)$, is $\mathcal O(m)$.
Crucially this number does not depend on either the data size, the dimension of the latent Gaussian $\theta$, nor the dimension of the hyperparameters $\phi$ and $\eta$.

\begin{figure}
  \begin{center}
    \includegraphics[width=3in]{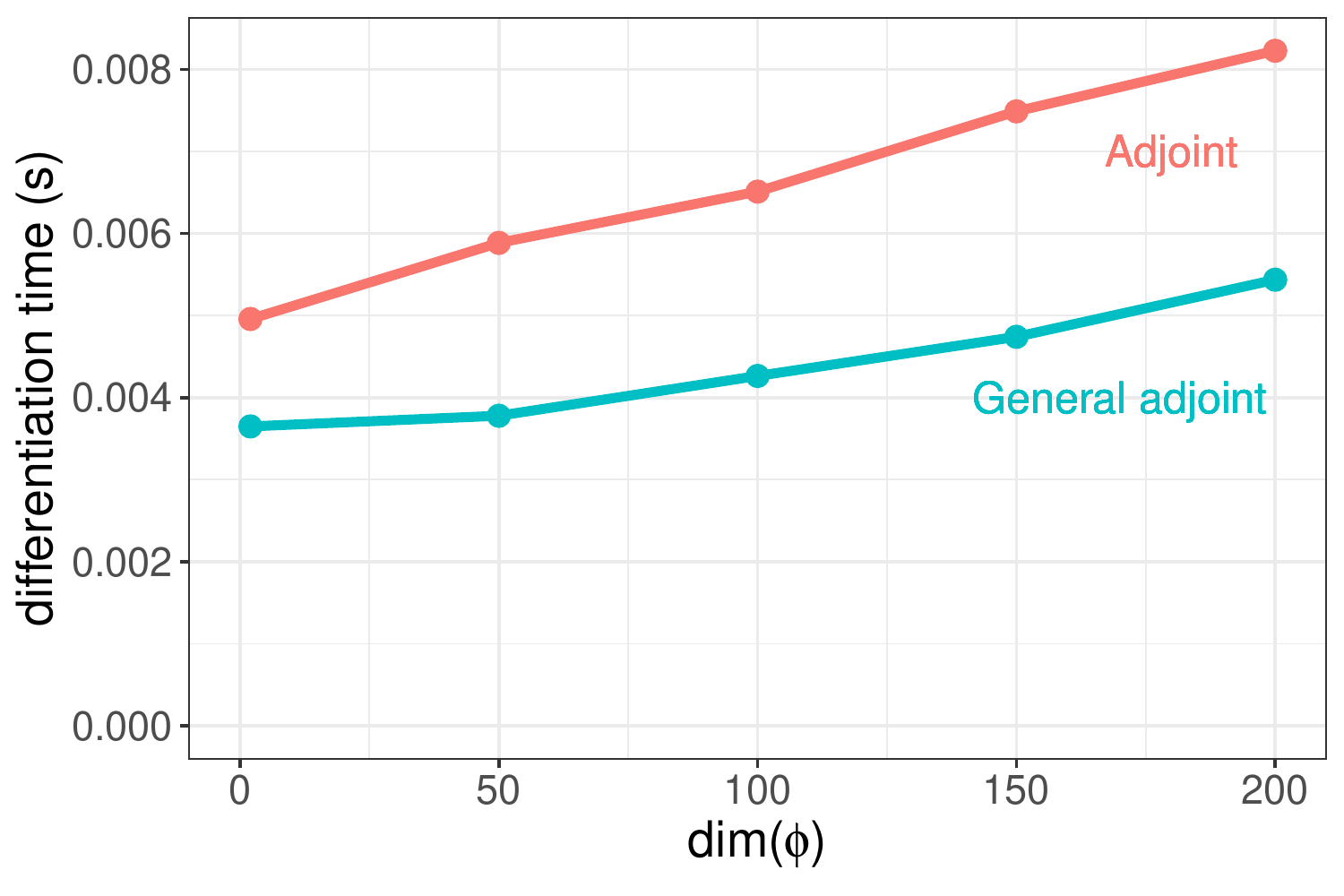}
  \end{center}
  \caption{Wall time to differentiate the marginal density of a SKIM using the general adjoint-differentiated Laplace approximation (Algorithm~\ref{algo:general-adjoint}),
  benchmarked against the method by \citet{Margossian:2020-laplace}.}
  \label{fig:compDiff2}
\end{figure}

The resulting algorithm only requires users to specify code to evaluate the prior covariance and the log likelihood but not their derivatives.
In general switching from analytical derivatives to automatic differentiation incurs some computational cost.
However adjoint methods, in addition to automating the computation of derivatives, bypass the evaluation of certain terms, which may be expansive to calculate even when analytical expressions exist.
When applied to the sparse kernel interaction model (SKIM) studied by \citet{Margossian:2020-laplace}, the general adjoint-differentiation outperforms the adjoint-differentiation, despite not having access to analytical derivatives (Figure~\ref{fig:compDiff2}).
Hence the generalization presents code which is more readable, more expandable, and slightly more efficient.

Understanding the benefits of the general adjoint-differentiated Laplace approximation on unconventional likelihoods remains ongoing work.
Section~\ref{sec:gam-experiment} examines a population pharmacokinetic model,
with a likelihood parameterized by a linear ordinary differential equation.
This examples demonstrates the integrated Laplace approximation can be very accurate in an unorthodox setting,
but highlights how challenging it can be to compute the approximation when the likelihood is not log-concave.

\section{Newton solvers and $B$-matrices} \label{sec:newton-solvers}

\begin{algorithm}
\caption{Abstract Newton solver} \label{algo:newton-abstract}
  {\bf input:} $y$, $\phi$, $\eta$,  $K$, $\pi(y \mid \theta, \eta)$, $\theta_0$, $\Delta$  \;
  $\theta \gets \theta_0$ \Comment initial guess \\
  $\Psi_\text{new} \gets -\infty$ \Comment initial objective \\
   \While {$(|\Psi_\mathrm{new} - \Psi_\mathrm{old}| \ge \Delta)$}{
   $W \gets - \nabla^2_\theta \log \pi(y \mid \theta, \eta)$ \\
   $\theta \gets (K^{-1} + W)^{-1} (W \theta + \nabla_\theta \log \pi(y \mid \theta, \eta))$ \\
   $\Psi_\text{old} \gets \Psi_\text{new}$ \\
   $\Psi_\text{new} \gets - \frac{1}{2} \theta^T K^{-1} \theta + \log \pi(y \mid \theta, \eta)$
  }
 \textbf{return:} $\log \pi_\mathcal{G}(y \mid \phi, \eta) = \Psi_\text{new} - \frac{1}{2} \log |K| |K^{-1} + W|$
\end{algorithm} 

Algorithm~\ref{algo:newton-abstract} describes an \textit{abstract Newton solver},
without the details required to insure a numerically stable implementation.
The expression for the approximate marginal likelihood, $\log \pi_\mathcal{G} (y \mid \phi, \eta)$,
is obtained via a Gaussian integral; see \citet[Section 3.4.4]{Rasmussen:2006}.
We use this abstraction to define a class of Newton solvers which can be used to compute the approximate marginal likelihood and its gradient,
while requiring minimal adjustments to our algorithm when we switch optimizer.

The main difficulty with a ``brute force'' implementation of the above solver is that we cannot safely invert the prior covariance matrix, $K$, or the sum $(K^{-1} + W)$, whose eigenvalues may be arbitrarily close to 0.
We will encounter similar issues when trying to invert $W$ in our calculation of the gradient.
Our main asset to avoid these difficult inversions is the Woodburry-Sherman-Morrison formula,
stated below for convenience.
\begin{lemma} \label{lemma:WSM}
(Woodburry-Sherman-Morrison formula.)
Given $Z \in \mathbb R^{n \times n}$, $W \in \mathbb R^{m \times m}$, $U \in \mathbb R^{n \times m}$, and $V \in \mathbb R^{n \times m}$,
we have
\begin{equation*}
  (Z + UWV^T)^{-1} = Z^{-1} - Z^{-1}U(W^{-1} + V^TZ^{-1}U)^{-1} V^T Z^{-1},
\end{equation*} 
where we assume the relevant inverses all exist.
\end{lemma}
%
%
Lemma~\ref{lemma:WSM} offers several decompositions we can take advantage of.
We consider three:
\begin{eqnarray} \label{eq:B-matrix}
(K^{-1} + W)^{-1} & = & K - K W^\frac{1}{2} (I + W^\frac{1}{2} K W^\frac{1}{2})^{-1} W^\frac{1}{2} K, \nonumber \\
  & = & K^\frac{1}{2} (I + K^{\frac{1}{2} T} W K^\frac{1}{2} )^{-1} K^{\frac{1}{2}T}, \nonumber \\
  & = & K - K W (I + KW)^{-1} K,
\end{eqnarray}
where $A^\frac{1}{2}$ is an equivalence class of matrices, such that $A^\frac{1}{2} A^{\frac{1}{2} T} = A$.

Overloading notation, I denote $B$ the matrix to invert on the RHS.
Remarkably all $B$-matrices have the same log determinant, 
which is equal to the log determinant needed to compute the approximate marginal distribution.
This can be shown using the determinant version of the Woodbury-Sherman-Morrison formula,
stated below for convenience.

\begin{lemma} 
(Woodbury-Sherman-Morrison for determinants)
  Given $Z \in \mathbb R^{n \times n}$, $W \in \mathbb R^{m \times m}$, $U \in \mathbb R^{n \times m}$, and $V \in \mathbb R^{n \times m}$,
we have
  \begin{equation*}
    |Z + UWV^T| = |Z| |W| |W^{-1} + V^TZ^{-1}U|,
  \end{equation*}
  where we assume the relevant inverses all exist.
\end{lemma}
Then some careful manipulations give us
\begin{eqnarray}
  \log |K| |K^{-1} + W| & = & \log |I + W^\frac{1}{2} K W^\frac{1}{2}| \nonumber \\
    & = & \log |I + K^{\frac{1}{2} T} W K^\frac{1}{2}| \nonumber \\
    & = & \log |I + KW|.
\end{eqnarray}

\subsection{$B = I + W^\frac{1}{2} K W^\frac{1}{2}$}

To use the first decomposition we need to compute $W^\frac{1}{2}$.
One option is to take the matrix square-root which requires $W$ to be 
quasi-triangular.\footnote{See the implementation in Eigen, \url{https://eigen.tuxfamily.org/dox/unsupported/group\_\_MatrixFunctions\_\_Module.html}.}
To achieve better computation we may exploit the sparsity of the Hessian, which is often block-diagonal.
Denoting $m \times m$ the size of each block and $n \times n$ the size of the Hessian,
computing the square root or the Cholesky decomposition block-wise costs $\mathcal O(m^2 n)$ operations, rather than $\mathcal O(n^3)$.
In the case where $W$ is positive-definite and diagonal, it suffices to take the element-wise square-root for a total complexity $\mathcal O(n)$.

The advantage of this decomposition is that $B$ is a symmetric matrix.
We can therefore compute a Cholesky decomposition, $L$,
and in turn use it for \textit{solve} methods, and to compute the log determinant of $B$,
\begin{eqnarray*}
  \log |B| = \log |L| |L^T| = 2 \sum_i \log L_{ii}.
\end{eqnarray*}

We can now write a numerically stable version of the Newton solver
under the assumption that $W^\frac{1}{2}$ exists (Algorithm~\ref{algo:newton-B1}).
This is the Newton solver used by \citet{Rasmussen:2006} and \citet{Margossian:2020-laplace},
except that we do not assume $W$ is diagonal.
One particularly useful detail is that at each Newton step we compute
\begin{equation}
  {\bf a} = K^{-1} \theta,
\end{equation}
without actually inverting $K$.
The vector $\bf a$ is then used to compute the objective function
and during the adjoint-differentiation step (Section~\ref{sec:g-adjoint-differentiation}).

The Newton iteration can be augmented with a linesearch step.
Let ${\bf a}_\text{old}$ and $\theta_\text{old}$ be respectively the $a$-vector and the guess for $\theta$ computed at the previous iteration.
Then reducing the step length by a factor of 2 is done by updating $\bf a$, given that
\begin{equation*}
  {\bf a} \gets \frac{ {\bf a} + {\bf a}_\text{old}}{2} \iff \theta \gets \frac{\theta_\text{new} + \theta_\text{old}}{2}.
\end{equation*}
This procedure is repeated until a chosen condition is met.
We may for example require the objective function, $\Psi$, to decrease at each Newton iteration.
Checking this condition is cheap, given that equipped with $\bf a$, the computation of the objective function is inexpensive.

If all our optimizers subscribe to a similar structure,
we can write algorithms for the approximate marginal likelihood and its gradient
which are (mostly) agnostic to which optimizer we use.

\begin{algorithm}[!h]
    \caption{Newton solver using $B = I + W^\frac{1}{2} K W^\frac{1}{2}$}
    \label{algo:newton-B1}
      \textbf{input:} $K$, $y$, $\pi(y \mid \theta, \phi)$, $\Delta$ \\
      $\theta \gets \theta_0$ \hspace{1cm} (initial guess) \\
      $\Psi_\text{new} \gets -\infty$ \ \ (initial objective) \\
      \While {$(|\Psi_\mathrm{new} - \Psi_\mathrm{old}| \ge \Delta)$}{
      $W \gets - \nabla_\theta \nabla_\theta \log \pi(y \mid \theta, \phi)$ \\
      $L \gets \mathrm{Cholesky}(I + W^\frac{1}{2} K W^\frac{1}{2})$  \\
      ${\bf b} \gets W \theta + \nabla_\theta \log \pi(y \mid \theta, \phi)$  \\
      ${\bf a} \gets {\bf b} - W^\frac{1}{2} L^T \setminus (L \setminus (W^\frac{1}{2} K {\bf b}))$  \\
      $ \theta \gets K {\bf a}$ \\
      $ \Psi_\text{old} = \Psi_\text{new}$ \\
      $ \Psi_\text{new} \gets -\frac{1}{2} {\bf a} \theta + \log \pi(y \mid \theta, \eta)$ \\
      }
      $\log \pi(y \mid \phi) \gets \Psi_\text{new} - \sum_i \log L_{ii}$ \\
      \textbf{return:} $\theta$, $\log \pi_\mathcal{G}(y \mid \phi)$
  \end{algorithm}


\subsection{$B = I + K^{\frac{1}{2} T} W K^\frac{1}{2}$}

This decomposition is proposed by \citet{Vanhatalo:2021} for the case where the likelihood is not log-concave meaning that $W$ is not amiable to any straightforward decomposition.
On the other hand, we assume $K$ can be safely inverted.
Since $K$ is a covariance matrix, $K^\frac{1}{2}$ can be computed using a Cholesky decomposition.
$B$ is still symmetric and admits a Cholesky decomposition, $LL^T = B$.

With this $B$-matrix, the Newton step is
\begin{eqnarray*}
  \theta^\text{new} & = & (K^{-1} + W)^{-1} (\nabla \log \pi(y \mid \theta, \eta) + W \theta) \\
    & = & K^\frac{1}{2} (I + K^{\frac{1}{2} T} W K^\frac{1}{2} )^{-1} K^\frac{1}{2} (\nabla \log \pi(y \mid \theta, \eta) + W \theta) \\
    & \triangleq & K^\frac{1}{2} {\bf c},
\end{eqnarray*}
which does not quite have the desired form $\theta^\text{new} = K {\bf a}$.
We can however compute ${\bf a} = K^{-\frac{1}{2}} {\bf c}$, which is cheap given $K^\frac{1}{2}$ is triangular.
Reconstructing $\bf a$ is not strictly necessary but it makes this approach more consistent with the Newton steps used for other $B$-matrices.

\subsection{$B = I + KW$}

The third and final decomposition makes no strong assumptions on $K$ and $W$.
The major drawback of this approach is that the $B$-matrix is not symmetric and therefore does not admit a Cholesky decomposition.
Instead, we resort to the more expansive LU-decomposition (with partial-pivoting given $B$ is invertible),
\begin{equation*}
  B = LU,
\end{equation*}
where $L$ is lower-triangle, $U$ upper-triangle, and which can then be used for solve methods with $B$
and to compute $\log |B|$.

Algorithm~\ref{algo:Newton-all} provides a general computation of the Laplace approximation,
which admits a choice of $B$-matrix as an option. 
\begin{algorithm}
\caption{General Newton solver. \textit{Writing all three Newton solvers in one algorithm highlights common features between the methods, which leads to lighter code and more general methods for calculating gradients.}} \label{algo:Newton-all}
  {\bf input:} $y$, $\phi$, $\eta$,  $K(\phi)$, $\pi(y \mid \theta, \eta)$, $B$-matrix \\
  $\theta = \theta_0$ \ \ (initial guess) \\
  $\Psi_\text{old} = - \infty$ \ \ (initial objective) \\
  \While{$(|\Psi_\mathrm{new} - \Psi_\mathrm{old}| > \Delta)$}{
    $W = - \nabla^2_\theta \log \pi(y \mid \theta, \eta)$ \\
    ${\bf b} = W \theta + \nabla_\theta \log \pi(y \mid \theta, \eta)$ \\
    \uIf {$(B = I + W^\frac{1}{2} K W^\frac{1}{2})$} {
      $L = \mathrm{Cholesky}(B)$ \\
      ${\bf a}  = {\bf b} - W^\frac{1}{2} L^T \backslash (L \backslash W^\frac{1}{2} K {\bf b}))$ \\
    }
    \uElseIf{$(B = I + K^{\frac{1}{2} T} W K^\frac{1}{2})$}{
      $L = \text{Cholesky}(B)$ \\
      ${\bf c} = L^T \backslash (L \backslash (K^\frac{1}{2} {\bf b}))$ \\
      ${\bf a} = K^\frac{1}{2} \backslash {\bf c}$ \\
    }
    \uElseIf{$(B = I + KW)$}{
        $(L, U) = \mathrm{LUdecomposition}(I + KW)$ \\
        ${\bf a} = {\bf b} - W U \backslash (L \backslash (K {\bf b}))$  \\
     }
    \textbf{end} \\
    $\Psi_\text{old} = \Psi_\text{new}$  \\
    $\Psi_\text{new} = - \frac{1}{2} {\bf a} \theta + \log \pi(y \mid \theta, \eta)$ \\
  \uIf{$(B = I + W^\frac{1}{2} K W^\frac{1}{2}$ or $B = I + K^{\frac{1}{2} T} W K^\frac{1}{2})$}{
    $\log|B| = 2 \sum_i \log L_{ii}$ \\
  }
  \uElseIf{$(B = I + KW)$}{
  $\log|B| = \sum_i \log L_{ii} + \sum_i \log U_{ii}$ \\
  }
  \textbf{end} \\
}
 \textbf{return:} $\log \pi_\mathcal{G}(y \mid \phi, \eta) = \Psi_\text{new} - \frac{1}{2} \log |B|$
\end{algorithm}

\section{Gradients with respect to $\phi$ and $\eta$}

We now derive expressions for the derivative of the approximate marginal likelihood.

\begin{proposition} \label{lemma:diff-phi}
 Without loss of generality, assume $K$ only depends on $\phi$, while the likelihood $\log \pi(y \mid \theta, \eta)$ only depends on $\eta$.
 Denote $\hat \theta$ the argument which maximizes $\pi(\theta \mid y, \phi, \eta)$.
 The derivative of the approximate marginal likelihood with respect to an element $\phi_j$ of $\phi$ is
 \begin{eqnarray}
   \frac{\partial}{\partial \phi_j} \log \pi_\mathcal{G}(y \mid \phi, \eta) & = & - \frac{1}{2} \hat \theta^T K^{-1} \frac{\partial K}{\partial \phi_j} K^{-1} \hat \theta \nonumber  \\
   & & - \frac{1}{2} \mathrm{trace} \left ( (W^{-1} + K)^{-1} \frac{\partial K}{\partial \phi_j} \right) \nonumber \\
   & & + \sum_{i = 1}^n \frac{\partial}{\partial \hat \theta_i} \log \pi_\mathcal{G} (y \mid \phi, \eta) \cdot (I + KW)^{-1} \frac{\partial K}{\partial \phi_j},
 \end{eqnarray}
 where we assume the requisite derivatives exist and
 \begin{equation}
  \frac{\partial}{\partial \hat \theta_i}  \log \pi_\mathcal{G}(y \mid \phi, \eta) = - \frac{1}{2} \mathrm{trace} \left ((K^{-1} + W)^{-1} \frac{\partial W}{\partial \hat \theta_i} \right).
\end{equation}
\end{proposition}
This result is worked out by \citet[Section 5.5.1]{Rasmussen:2006}.
A similar result is obtained when differentiating with respect to $\eta$.

\begin{proposition} \label{lemma:diff-eta}
  The derivative of the approximate marginal likelihood with respect to an element $\eta_l$ of $\eta$ is
  \begin{eqnarray} \label{eq:gradient-eta}
  \frac{\partial}{\partial \eta_l}\log \pi_\mathcal{G} (y \mid \phi, \eta) & = &  \frac{\partial }{\partial \eta_l} \log \pi(y \mid \hat \theta, \eta) \nonumber \\
  & & + \frac{1}{2} \mathrm{trace} \left ((K^{-1} + W)^{-1} \frac{\partial \nabla^2_\theta \log \pi(y \mid \hat \theta, \eta)}{\partial \eta_j} \right) \nonumber \\
 & & + \sum_{i = 1}^n \frac{\partial \log \pi_\mathcal{G} (y \mid \phi, \eta)}{\partial \hat \theta_i} \cdot (I + KW)^{-1} K \frac{\partial}{\partial \eta_l} \nabla_\theta \log \pi(y \mid \hat \theta, \eta),
\end{eqnarray}
where we assume the requisite derivatives exist.
\end{proposition}
The proof is in the Appendix at the end of this chapter.
Several of the above expressions can be simplified in the special cases where $W$ or $K$ are diagonal.

\begin{remark} \label{remark:diff-likelihood}
The derivative with respect to either $\eta$ or $\theta$ decomposes into three terms:
\begin{itemize}
  \item[] (i) an explicit term (partial derivative of the objective function),
  \item[]
  \item[] (ii) the derivative of a log determinant, which becomes a trace,
  \item[]
  \item[] (iii) a dot product with a gradient with respect to $\hat \theta$.
\end{itemize}
The expressions in Propositions~\ref{lemma:diff-phi} and \ref{lemma:diff-eta} are organized accordingly.
I will use this organization in Section~\ref{sec:diff-marginal}. 
\end{remark}

Once again we must contend with inverted matrices, namely $(K^{-1} + W)^{-1}$,
which we have already dealt with when building the Newtons solver,
and $(I + KW)^{-1}$.
It turns out the latter can also be handled with decompositions of the $B$-matrix performed during the final Newton step, meaning no further Cholesky or LU decomposition is required.
Indeed
\begin{equation} \label{eq:R-decomposition}
  (I + KW)^{-1} = I - K(K + W^{-1})^{-1},
\end{equation}
and $(K + W^{-1})^{-1}$ can be expressed in terms of any of three $B$-matrices we are working with:
\begin{eqnarray}
  R \triangleq (K + W^{-1})^{-1} & = & W^\frac{1}{2} (I + W^\frac{1}{2} K W^\frac{1}{2})^{-1} W^\frac{1}{2} \nonumber \\
    & = & W - W K^\frac{1}{2} (I + K^{\frac{1}{2} T} W K^\frac{1}{2})^{-1} K^{\frac{1}{2} T} W \nonumber \\
    & = & W - W(I + KW)^{-1} KW.
\end{eqnarray}
The last equality is superfluous, since we can directly handle the original matrix $(I + KW)^{-1}$
if we use $B = I + KW$.
To run the adjoint-differentiated Laplace approximation by \citet{Margossian:2020-laplace},
extended to handle derivatives with respect to $\eta$,
methods to compute the following derivatives (analytically or otherwise) need to be provided:
\begin{itemize}
 \item $\nabla_\theta \log \pi(y \mid \theta, \eta)$,
 \item $\nabla^2_\theta \log \pi(y \mid \theta, \eta)$,
 \item $\nabla^3_\theta \log \pi(y \mid \theta, \eta)$,
 \item $\nabla_\eta \log \pi(y \mid \theta, \eta)$,
 \item $\nabla_\eta \nabla_\theta \log \pi(y \mid \theta, \eta)$,
 \item $\nabla_\eta \nabla^2_\theta \log \pi(y \mid \theta, \eta)$.
\end{itemize}
 The requirement to calculate higher-order derivatives of the likelihood makes the implementation of the integrated Laplace approximation cumbersome.
 Our next task is therefore to relax this requirement.

\section{Automatic differentiation of the likelihood}

Fortunately we can eliminate the burden of analytically computing derivatives by applying automatic differentiation, as was already done to remove calculations of $\partial K / \partial \phi$.

\subsection{Allowed operations with automatic differentiation}

Consider a function
\begin{eqnarray*}
  f : & \mathbb R^m \to \mathbb R^n \\
     & x \to f(x).
\end{eqnarray*}
A forward mode sweep of automatic differentiation allows us to compute the directional derivative
\begin{equation*}
  \frac{\partial f}{\partial x} \cdot v
\end{equation*}
for an initial tangent $v \in \mathbb R^m$.
A reverse mode sweep, on the other hand, computes the co-directional derivative
\begin{equation*}
w^T \cdot \frac{\partial f}{\partial x}
\end{equation*}
for an initial cotangent $w \in \mathbb R^n$.
We can compute higher-order derivatives by iteratively applying sweeps of automatic differentiation.
This procedure is straightforward for forward mode, less so for reverse mode.
When doing multiple sweeps in \texttt{Stan}, we may only use a single reverse mode sweep
and this sweep must be the final sweep \citep{Carpenter:2015}.

\subsection{Principles for an efficient implementation}

I observe the following principles to write an efficient implementation:
\begin{enumerate}
  \item \textit{Contraction.} Avoid computing full Jacobian matrices. Instead, only compute directional derivatives by contracting Jacobian matrices with the right tangent or cotangent vectors.
  \item[]
  \item \textit{Linearization.} In the original algorithm, identify linear operators, $\Phi$, which take in and return a derivative, e.g.
  \begin{equation*}
    \frac{\partial c}{\partial a_i} = \Phi \left (\frac{\partial b}{\partial a_i} \right) \in \mathbb R
  \end{equation*}
  Then rather than compute the derivatives one element at a time,
  compute $\Phi(b)$ and apply a reverse mode sweep of automatic differentiation
  to obtain the desired gradient.
  This can be seen as a strategy to identify the directions along which to compute derivatives.
  \item[]
  \item \textit{Sparse computation.} For sparse objects of derivatives, only compute the non-zero elements,
  again by carefully picking the directions along which we compute the derivatives.
\end{enumerate}
At each step of the Newton solver, I compute the full negative Hessian, $W$, even though this is not strictly necessary.
This is because (i) the Hessian is typically sparse and therefore relatively cheap to compute,
and (ii) $W$ is used many times both in the computation of the Laplace approximation and its differentiation.

\subsection{Differentiating the negative Hessian, $W$}

Most LGMs admit a likelihood with a block-diagonal or even diagonal Hessian.
In typical automatic differentiation frameworks, the cost of computing a block-diagonal Hessian with block size $m \times m$ is $2m$ sweeps.
This is notably the case with \texttt{Stan} as I will demonstrate.
Somewhat contrary to general wisdom, it is actually possible to compute a Hessian matrix with a single reverse mode sweep using a graphical model and an algebraic model \citep{Gower:2011}.
I have not investigated this approach but believe it is promising.\footnote{... following a conversation with Robert Gower.}

\subsubsection{Diagonal Hessian.}

Consider a function
\begin{eqnarray*}
  f : & \mathbb R^n & \to \mathbb R\\
      & \theta & \to f(\theta).
\end{eqnarray*}
Suppose for starters that $f$ admits a diagonal Hessian.
To get the non-zero elements of $\nabla^2_\theta \log \pi(y \mid \theta, \eta)$, we only need to compute the Hessian-vector product,
\begin{equation*}
  [ \nabla^2_\theta \log \pi(y \mid \theta, \eta) ] \cdot \bf v,
\end{equation*}
where ${\bf v} = (1, 1, ..., 1) \in \mathbb R^n$.
This operation can be done using one forward mode and one reverse mode sweep of automatic differentiation.
In details: we introduce the auxiliary function
\begin{eqnarray*}
  g :& \mathbb R^n & \to \mathbb R, \\
      & \theta & \to \nabla_\theta \log \pi(y \mid \theta, \eta) \cdot {\bf v},
\end{eqnarray*}
which returns the sum of the partial derivatives and can be computed using one forward sweep.
Then ${\bf a}^T [\nabla_\theta g(\theta)]$ can be computed using one reverse mode sweep, where ${\bf a} = (1)$ is a vector of length 1 which contracts the $1 \times n$ gradient into a scalar.
Concisely
\begin{equation}
  [ \nabla^2_\theta \log \pi(y \mid \theta, \eta) ] \cdot {\bf v} = {\bf a}^T \cdot \nabla_\theta (\nabla_\theta \log \pi(y \mid \theta, \eta) \cdot {\bf v}).
\end{equation}

A similar procedure can be used to compute the diagonal tensor of third-order derivatives,
noting that it too only contains $n$ non-zero elements:
\begin{equation}
  \text{Diag} \left (\nabla^3_\theta \log \pi(y \mid \theta, \eta) \right) = {\bf a}^T \cdot \nabla_\theta \left( \nabla_\theta \left (\nabla_\theta f \cdot {\bf v} \right) \cdot {\bf v} \right). 
\end{equation}
This operation is performed in three sweeps, starting from the inner-most parenthesis and expanding out.

\subsubsection{Block-diagonal Hessian.}

Now suppose $f$ admits a $2 \times 2$ block diagonal Hessian.
I use the $2 \times 2$ case to develop some intuition before generalizing.
Consider the vectors ${\bf v_1} = (1, 0, 1, 0, ..., 1, 0) \in \mathbb R^n$ and ${\bf v_2} = (0, 1, ..., 0, 1) \in \mathbb R^n$.
Then
\begin{equation*}
  \nabla^2 f \cdot {\bf v_1} = \begin{pmatrix} \partial^2_{\theta_1^2} f & \partial^2_{\theta_1 \theta_2} f & 0 & 0 & ... \\ 
   \partial^2_{\theta_1 \theta_2} f & \partial^2_{\theta_2^2} f  & 0 & 0 & ... \\
   0 & 0 & \partial^2_{\theta_3^2} f & \partial^2_{\theta_3 \theta_4} f & ... \\
   0 & 0 & \partial^2_{\theta_3 \theta_4} f & \partial^2_{\theta_4^2} f & ... \\
   . . . & ... & ... & ... & ...
   \end{pmatrix}
   \begin{pmatrix} 1 \\ 0 \\ 1 \\ 0 \\ ... \end{pmatrix}
   = \begin{pmatrix} \partial^2_{\theta_1^2} f \\
         \partial^2_{\theta_1 \theta_2} f \\
         \partial^2_{\theta_3^2} f \\
         \partial^2_{\theta_3 \theta_4} f \\
         ...
      \end{pmatrix}
\end{equation*}
and
\begin{equation*}
  \nabla^2 f \cdot {\bf v_2} = \begin{pmatrix} \partial^2_{\theta_1^2} f & \partial^2_{\theta_1 \theta_2} f & 0 & 0 & ... \\ 
   \partial^2_{\theta_1 \theta_2} f & \partial^2_{\theta_2^2} f  & 0 & 0 & ... \\
   0 & 0 & \partial^2_{\theta_3^2} f & \partial^2_{\theta_3 \theta_4} f & ... \\
   0 & 0 & \partial^2_{\theta_3 \theta_4} f & \partial^2_{\theta_4^2} f & ... \\
   . . . & ... & ... & ... & ...
   \end{pmatrix}
   \begin{pmatrix} 0 \\ 1 \\ 0 \\ 1 \\ ... \end{pmatrix}
   = \begin{pmatrix} \partial^2_{\theta_1 \theta_2} f \\
         \partial^2_{\theta_2^2} f \\
         \partial^2_{\theta_3 \theta_4} f \\
         \partial^2_{\theta_4^2} f \\
         ...
      \end{pmatrix}.
\end{equation*}
These two Hessian-vector products return all the non-zero elements of the Hessian.
Thus computing the full Hessian requires 2 times the effort to evaluate a diagonal Hessian.
Specifically, we first compute $\nabla f \cdot {\bf v_1}$ using a forward sweep, followed by one reverse mode sweep, and repeat the process with ${\bf v_2}$.
The total cost for computing a $2\times2$ block-diagonal Hessian is thus 4 sweeps.\footnote{If we could start with reverse mode and then run forward mode, we could imagine computing the Hessian in 3 sweeps.}
The average cost of the sweeps can be reduced by exploiting the symmetry of the Hessian \citep[e.g][]{Griewank:2008, Gower:2011}.
 
In the $m \times m$ block-diagonal case, we need to construct $m$ initial tangents,
\begin{eqnarray} \label{eq:init-tangents-v}
  \bf v_1 & = (\underbrace{1, 0, \cdots, 0}_m, \underbrace{1, 0, \cdots, 0}_m, \cdots, \underbrace{1, 0, \cdots, 0}_m) \nonumber \\
  \bf v_2 & = (\underbrace{0, 1, \cdots, 0}_m, \underbrace{0, 1, \cdots, 0}_m, \cdots, \underbrace{0, 1, \cdots, 0}_m) \nonumber \\
  \vdots \nonumber \\
  \bf v_{m} & = (\underbrace{0, 0, \cdots, 1}_m, \underbrace{0, 0, \cdots, 1}_m, \cdots, \underbrace{0, 0, \cdots, 1}_m).
\end{eqnarray}
The elements of the Hessian are then computed using one forward mode along $m$ directions,
followed by a reverse mode.

\subsection{Differentiating the approximate marginal density} \label{sec:diff-marginal}

Differentiating the approximate marginal density with respect to the hyperparameters $\phi$ and $\eta$ requires computing three terms: (i) an explicit term, (ii) a differentiated log determinant, and (iii) a dot product with a derivative with respect to $\theta$ (Lemmas~\ref{lemma:diff-phi} and \ref{lemma:diff-eta}, and Remark~\ref{remark:diff-likelihood}).
The terms for $\phi$ are handled by \citet{Rasmussen:2006} and \citet{Margossian:2020-laplace},
so I will focus on $\eta$.
The explicit term for $\eta$ is simply the partial derivative of $\log \pi(y \mid \theta, \eta)$ with respect to $\eta$ which can be obtained using one reverse mode sweep of automatic differentiation.

\subsubsection{Differentiating the log determinant.}

Differentiating a log determinant produces a trace.
From Lemmas~\ref{lemma:diff-phi} and \ref{lemma:diff-eta} we see that we must evaluate a higher-order derivative inside a trace with respect to both $\eta$ and $\hat \theta$,
\begin{equation*}
  \text{trace} \left ( (K^{-1} + W)^{-1} \frac{\partial W}{\partial \eta_l} \right)
  \ \ \ \text{and} \ \ \ \text{trace} \left ( (K^{-1} + W)^{-1} \frac{\partial W}{\partial \hat \theta_j} \right).
\end{equation*}
Here I apply the linearization principle of automatic differentiation.
Let $A = (K^{-1} + W)^{-1}$ and note that trace$(AW)$ is a linear function of $W$.
Once we \textit{evaluate} and \textit{tape} $\text{trace}(AW)$,
it remains to do one reverse mode sweep to obtain a gradient with respect to $\eta$ and $\hat \theta$.
I use the term \textit{tape} to indicate that we must store the expression graph for trace$(AW)$ in order to perform the reverse mode sweep, given $W$ depends on both $\eta$ and $\hat \theta$.

Now
\begin{equation}
  \text{trace}(AW) = \sum_i \sum_k A_{ik} W_{ki} = \sum_i \sum_k A_{ik} W_{ik},
\end{equation}
where the second equality follows from the fact $W$ is symmetric.
Without loss of generality, assume $W$ is block diagonal, with block size $m \times m$.
As in our calculations of the Hessian, we start with a forward mode sweep along an initial tangent $v_1$. 
Next we do a second forward mode sweep with an initial tangent which contains the first column of each $m \times m$ block in A,
\begin{equation}  \label{eq:init-tangents-w}
  {\color{orange} {\bf w}_1} = (A_{1,1}, A_{2,1}, \cdots, A_{m, 1}, A_{m + 1, m + 1}, A_{m + 2, m + 1}, \cdots).
\end{equation}
That is ${\bf w}_1$ contains the colored elements in the below representation of $A$,
\begin{equation*}
  \begin{bmatrix}
  {\color{orange} A_{1, 1}} & A_{1, 2} & \cdots & \cdots & \cdots & \cdots  \\
  {\color{orange} A_{2, 1}} & A_{2, 2} & \cdots & \cdots & \cdots & \cdots \\
  \vdots & \vdots & \ddots & \ddots & \ddots & \ddots  \\
  {\color{orange} A_{m, 1}} & A_{m, 2} & \cdots & \cdots & \cdots & \cdots \\
  A_{m + 1, 1} & \cdots & \cdots & \color{orange} A_{m + 1, m + 1} & A_{m + 1, m + 2} & \cdots \\
  \cdots & \cdots & \cdots & \color{orange} A_{m + 2, m + 1} & A_{m + 2, m + 2} & \cdots \\
  \vdots & \vdots & \vdots & \vdots & \vdots & \ddots \\
  A_{2m, 1} & \cdots & \cdots & \color{orange} A_{2m, m + 1} & A_{2m, m + 2} & \cdots \\
  A_{2m + 1, 1} & \cdots & \cdots & A_{2m + 1, m + 1} & A_{2m + 1, m + 2} & \cdots \\
  \vdots & \vdots & \vdots & \vdots  & \vdots & \ddots
  \end{bmatrix}.
\end{equation*}
To obtain the full trace, we repeat this process $m$ times using the appropriate initial tangents, ${\bf v}_j$ and ${\bf w}_j$, $j \in [m]$.
It may be surprising that computing a scalar would require so many sweeps but this seems to be because there is an implicit matrix-matrix multiplication, which prevents us from describing the sum given by the trace using only two tangent vectors.
In other words, we cannot describe all the relevant $nm$ elements of $A$ using only $2n$ elements,
unless $A$ is low-rank.

To avoid repeating these calculations for $\hat \theta$ and $\eta$,
I compute the forward mode sweeps with respect to both $\hat \theta$ and $\eta$ at once.
To do so, I append $T$ 0's to $v_j$ and $w_j$, where $T$ is the dimension of $\eta$, e.g.
\begin{eqnarray} \label{eq:init-tangents}
  {\bf v}^*_1 & = & (\underbrace{1, 0, \cdots, 0}_m, \underbrace{1, 0, \cdots, 0}_m, \cdots, \underbrace{0, 0, \cdots, 0}_T), \nonumber \\
  {\bf w}^*_1 & = &(\underbrace{A_{1,1}, A_{2,1}}_m, \cdots, \underbrace{A_{m, 1}, A_{m + 1, m + 1}, A_{m + 2, m + 1}}_m, \cdots, \underbrace{0, 0, \cdots, 0}_T).
\end{eqnarray}
Unfortunately this procedure requires computing $A$, a potentially expansive operation.
The cost can be reduced by only computing the block-diagonal elements of $A$.

\subsubsection{Differentiating the dot product.}

The dot product results from the chain rule, which requires us to multiply $\partial \log \pi_\mathcal{G} (y \mid \phi, \eta)  / \partial \hat \theta_j$
and $\partial \hat \theta_j / \partial \eta_l$.
Let
\begin{equation}
  s_2 \triangleq \nabla_{\hat \theta} \log \pi_\mathcal{G} (y \mid \phi, \eta) = - \frac{1}{2} \text{trace} \left ((K^{-1} + W)^{-1} \right).
\end{equation}
$s_2$ can be computed using the method outlined in the previous section.
Once again these calculations are subject to useful simplifications when $W$ is diagonal.

Per Proposition~\ref{lemma:diff-eta}, the dot product is taken with 
\begin{eqnarray}
  s_3 & \triangleq & (I + KW)^{-1} K \frac{\partial}{\partial \eta_l} \nabla_\theta \log \pi(y \mid \hat \theta, \eta) \nonumber \\
  & = & (I - KR) K \frac{\partial}{\partial \eta_l} \nabla_\theta \log \pi(y \mid \hat \theta, \eta),
\end{eqnarray}
where the second equality follows from Equation~\ref{eq:R-decomposition}
and $R$ can safely be computed using any of the three $B$-matrices introduced in Section~\ref{sec:newton-solvers}.
Here too I apply the linearization principle and consider the function
\begin{equation}
  \tilde s_3 \triangleq (I - KW)^{-1} K \nabla_{\hat \theta} \log \pi(y \mid \hat \theta, \eta) = (I - KR) K \nabla_{\hat \theta} \log \pi(y \mid \hat \theta, \eta).
\end{equation}
The scalar $s^T_2 \tilde s_3$ can be computed using one forward sweep with respect to $\hat \theta$ and initial tangent
\begin{equation} \label{eq:tangent-dot}
  {\bf u} = K (I - KW)^{-1} s_2 = K (I - KR) s_2,
\end{equation}
where I drop the transpose on the term before $s_2$ due to symmetry.
It then remains to apply a reverse mode sweep with respect to $\eta$.

\subsection{General adjoint-differentiation} \label{sec:g-adjoint-differentiation}

We are finally ready to write down the general adjoint-differentiation (Algorithm~\ref{algo:general-adjoint}).

Let
\begin{equation*}
  \text{AD}(f, [\text{fwd}, {\bf v}, x])
\end{equation*}
be a single forward mode sweep which returns $\partial f / \partial x \cdot {\bf v}$.
Similarly, $\text{AD}(f, [\text{rev}, {\bf w}, x])$ returns ${\bf w}^T \cdot \partial f / \partial x$.
Furthermore, we can encode multiple sweeps, for example $\mathrm{AD}(f, [\text{fwd}, {\bf v}, x], [\text{rev}, {\bf w}, x])$,
where sweeps are executed from the left to the right.

\begin{algorithm}
   \caption{General adjoint-differentiation. \textit{Note: additional specialized steps can be taken in the case where $W$ or $K$ is diagonal.}}
   \label{algo:general-adjoint}
    \textbf{input:} $y$, $\phi$, $\eta$, $\pi(y \mid \theta, \eta)$ \\
    \textbf{saved input from the  Newton solver}: $\hat \theta$, ${\bf a}$, $\nabla_{\hat \theta} \log \pi(y \mid \hat \theta, \eta)$, $K$, $B$-matrix \\
    \ \ $W^\frac{1}{2}$, $L$ \Comment{$B = I + W^\frac{1}{2} K W^\frac{1}{2}$, $LL^T = B$} \\
    \ \ $W$, $K^\frac{1}{2}$, $L$ \Comment{$B = I + K^{\frac{1}{2}T} W K^\frac{1}{2}$, $LL^T = B$} \\
    \ \ $W$, $L$, $U$ \Comment{$B = I + KW$, $LU = B$} \\
    \uIf{$(B =  I + W^\frac{1}{2} K W^\frac{1}{2})$}{
      $R \gets W^\frac{1}{2} L^T \backslash (L \backslash W^\frac{1}{2})$ \Comment $R = (K + W^{-1})^{-1}$\\
      $C \gets L \backslash (W^\frac{1}{2} K)$ \\
      $A \gets K - C^TC$ \Comment $A$ contains initial tangents for log det. derivative.
    }
    \uElseIf{$(B = I + K^{\frac{1}{2}T} W K^\frac{1}{2})$}{
      $D \gets L \backslash K^\frac{1}{2} W$ \\
      $R \gets W - D^TD$ \\
      $C \gets L \backslash K^{\frac{1}{2}T}$ \\
      $A \gets C^TC$
    }
    \uElseIf{$(B = I + KW)$}{
      $R \gets W - WCW$ \\
      $C \gets U \backslash L \backslash K$  \\
      $A \gets K - KWC$
    }
    \textbf{end} \\
    $V \gets [{\bf v}^*_1, {\bf v}^*_2, \cdots, {\bf v}^*_m]$  \Comment Initial tangents, Equations~\ref{eq:init-tangents-v} and \ref{eq:init-tangents} \\
    $W \gets [{\bf w}^*_1, {\bf w}^*_2, \cdots, {\bf w}^*_m]$ \Comment Initial tangents using $A$, Equations~\ref{eq:init-tangents-w} and \ref{eq:init-tangents} \\
    $s \gets 0$ \\
    \For{$i \in \{1, 2, \cdots, m\}$}{
      $s \gets s + \text{AD}(\log \pi(y \mid \hat \theta, \eta), [\text{fwd}, {\bf v}^*_i, \vartheta],
      [\text{fwd}, {\bf w}^*_i, \vartheta])$ \Comment $\vartheta \triangleq (\hat \theta, \eta)$
    }
    $s \gets \text{AD}(s, [\text{rev}, 1, \vartheta])$  \\
    $s_2 \gets s[1:n]$ \Comment Gradient of log determinant w.r.t $\hat \theta$ \\
    $s_2' \gets s[(n + 1):(n + T)]$ \Comment Gradient of log determinant w.r.t $\eta$ \\
    $\Omega^T = \frac{1}{2} {\bf a a}^T - \frac{1}{2} R + (s_2 + RKs_2)[\nabla{\hat \theta} \log \pi(y \mid \hat \theta, \eta)]^T$ \Comment \citep[3.2]{Margossian:2020-laplace}  \\
    $\nabla_\phi \log \pi_\mathcal{G} (y \mid \phi, \eta) = \text{AD}(K, [\text{rev}, \Omega, \phi])$ \\
    \uIf{$(B =  I + W^\frac{1}{2} K W^\frac{1}{2})$ or $(B = I + K^{\frac{1}{2}T} W K^\frac{1}{2})$}{
      ${\bf u} = K (I - KR) s_2$  \Comment Initial tangent, Equation~\ref{eq:tangent-dot}
    }
    \uElseIf{$(B = I + KW)$}{
      ${\bf u} = U \backslash L \backslash K$  \Comment Initial tangent, Equation~\ref{eq:tangent-dot}
    }
    {\bf end} \\
    $\nabla_\eta \log \pi_\mathcal{G}(y \mid \phi, \eta) = \text{AD}(\log \pi(y \mid \hat \theta, \eta), [\text{rev}, 1, \eta]) + s_2'$ \Comment Proposition~\ref{lemma:diff-eta} \\
    $ \hspace{1.5in}  + \text{AD}(\log \pi(y \mid \hat \theta, \eta), [\text{fwd}, \mathbf u, \theta], [\text{rev}, 1, \eta])$ \\
    \textbf{return:}  $\nabla_\phi \log \pi_\mathcal{G} (y \mid \phi, \eta)$, $\nabla_\eta \log \pi_\mathcal{G}(y \mid \phi, \eta)$
\end{algorithm}

\section{Posterior draws for the marginalized out parameters}

I review a procedure to generate posterior draws for the latent Gaussian variable, $\theta$,
following \citet{Rasmussen:2006}, and show how this approach fits in the $B$-matrix framework.

After generating posterior draws for the hyperparameters, $(\phi, \eta)$,
we recover posterior draws for $\theta$ using the Laplace approximation:
\begin{eqnarray*}
  (\phi, \eta) & \sim & \pi_\mathcal{G}(\phi, \eta \mid y), \\
  \theta & \sim & \pi_\mathcal{G} ( \theta \mid y, \phi, \eta).
\end{eqnarray*}
We may also want to draw new latent Gaussian variables, $\theta^*$,
given a prior distribution $\pi(\theta, \theta^*)$.
In a Gaussian process, the covariance matrix, $K$, is typically parameterized by $\phi$ and a covariate $X$.
Hence, for a new set of observations, the new prior covariance would be $K(\phi, X^*)$.
Let $\bf K$ be the joint covariance matrix over $(\theta, \theta^*)$,
\begin{equation*}
  {\bf K} = \begin{bmatrix}
  K(X, X) & K(X, X^*)\\
  K(X, X^*) & K(X^*, X^*)
  \end{bmatrix},
\end{equation*}
and for convenience let $K = K(X, X)$, $K^* = K(X, X^*) = K(X^*, X)$ and $K^{**} = K(X^*, X^*)$.
Then
\begin{equation*}
{\bf K} = \begin{bmatrix}
K & K^*\\
K^* & K^{**}
\end{bmatrix}.
\end{equation*}
%
%
The mean of the approximating normal is
\begin{eqnarray*}
  \mathbb E_\mathcal{G} (\theta^* \mid X, y, \phi, X^*) = K^*K^{-1} \hat \theta = K^* \nabla_{\hat \theta} \log \pi(y \mid \hat \theta, \eta).
\end{eqnarray*}
The second equality follows from the fact $\hat \theta$ is the mode of the conditional distribution, $\pi(\theta \mid y, \phi, \eta)$, meaning
\begin{equation*}
  \nabla \Psi = 0 \implies \hat \theta = K \nabla_{\hat \theta} \log \pi(y \mid \hat \theta, \eta).
\end{equation*}
%
Furthermore,
\begin{eqnarray*}
  \Sigma_\mathcal{G}(\theta^* \mid y, \phi, \eta, X, X^*) = K^{**} - K^* (K + W^{-1})^{-1} K^*.
\end{eqnarray*}
The procedure is summarized in Algorithm~\ref{algo:post-sample}.
Here the computation is dominated by the evaluation of the covariance matrix, $\Sigma_\mathcal{G}$.

\begin{algorithm}
  \caption{Posterior draws for latent Gaussian $\theta^*$}
  \label{algo:post-sample}
  {\bf intput:} $y, \phi, \eta, X, X^*, K(\phi, X, X^*), \pi(y \mid \theta, \eta)$ \\
  {\bf saved input from the Newton solver:} $\hat \theta$, $W$, $K$, $\nabla_{\hat \theta} \log \pi(y \mid \hat \theta, \eta)$  \\
    \ \ $W^\frac{1}{2}$, $L$ \Comment{$B = I + W^\frac{1}{2} K W^\frac{1}{2}$, $LL^T = B$} \\
    \ \ $W$, $K^\frac{1}{2}$, $L$ \Comment{$B = I + K^{\frac{1}{2}T} W K^\frac{1}{2}$, $LL^T = B$} \\
    \ \ $W$, $L$, $U$ \Comment{$B = I + KW$, $LU = B$} \\
   $K^* \gets K(X, X^*)$  \\
   $K^{**} \gets K(X^*, X^*)$ \\
   $\mu^* \gets K^* \nabla_{\hat \theta} \log \pi(y \mid \theta, \eta)$ \\
   \uIf{$(B = I + W^\frac{1}{2} K W^\frac{1}{2})$}{
     $V \gets L \backslash W^\frac{1}{2} K^*$ \\
     $\Sigma^* \gets K^{**} - V^T V$
   }
   \uElseIf{$(B = I + K^{\frac{1}{2}T} W K^\frac{1}{2})$}{
      $D \gets L \backslash K^\frac{1}{2} W$ \\
      $R \gets W - D^TD$ \\
      $\Sigma^* \gets K^{**} - K^*RK^*$ \\
   }
   \uElseIf{$(B = I + KW)$}{
     $\Sigma^* = K^{**} - K^* (W - W U \backslash L \backslash K W)K^*$
   }
   \textbf{end} \\
   $\theta^* \sim \text{Normal}(\mu^*, \Sigma^*)$ \\
   \textbf{return:} $\theta^*$.
\end{algorithm}

\section{Numerical experiment} \label{sec:gam-experiment}

The integrated Laplace approximation is a well established method with success in many applications; see \cite{Rue:2017} and references therein.
A handful of papers study the potential of the integrated Laplace approximation when combined with MCMC, which is useful to expand the range of priors we can use \citep[e.g.][]{Gomez:2018, Monnahan:2018, Margossian:2020-laplace}.
Applications to less conventional likelihoods include the work by \citet{Vanhatalo:2009, Jylanki:2011, Joensuu:2012, Riihimaki:2014, Vanhatalo:2021}.
I examine the application of Hamiltonian Monte Carlo (HMC) using a general adjoint-differentiated Laplace approximation to a population pharmacokinetic model.
This model falls outside the traditional framework of LGMs and is meant to stress-test our algorithm, while taking advantage of its flexibility.

 \begin{figure}
    \begin{center}
    \begin{tikzpicture}
    [
      Round/.style={circle, draw=black!, fill=green!0, thick, minimum size=15mm},
      Empty/.style={circle, draw=white!, fill=green!0, thick, minimum size=1mm},
    ]
    
    \node[Round] (gut) at (0, 0) {Gut};
    \node[Round] (central) at (3, 0) {Central};
    \node[Empty] (k1) at (1.5, 0.5) {$k_1$};
    \node[Empty] (k2) at (4.5, 0.5) {$k_2$};

    \path [->, draw=black] (gut) -- (central);
    \path [->, draw=black] (central) -- (5.5, 0);

  \end{tikzpicture}
  \end{center}
  \caption{One compartment model with first-order absorption from the gut. \textit{The drug enters the body through the gut (bolus dose) and is then absorb in to the central compartment (blood and tissues) at a rate $k_1$. Over time, the drug is cleared at a rate $k_2$.}}
  \label{fig:oneCpt}
  \end{figure}
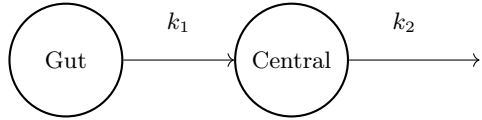

The one-compartment pharmacokinetic model with a first-order absorption from the gut describes the diffusion of an orally administered drug compound in the patient's body (Figure~\ref{fig:oneCpt}).
Two parameters of interest are the absorption rate, $k_1$, and the clearance rate, $k_2$.
The population model endows each patient with their own rate parameters,
subject to a hierarchical normal prior.
Of interest are the population parameters, $k_{1, \text{pop}}$ and $k_{2, \text{pop}}$,
and their corresponding population standard deviation, $\tau_{1, \text{pop}}$ and $\tau_{2, \text{pop}}$.

The full model is given below
\begin{eqnarray*}
  \text{\textit{hyperpriors}} \\
  k_{1, \text{pop}} & \sim & \text{Normal}(2, 0.5) \\
  k_{2, \text{pop}} & \sim & \text{Normal}(1, 0.5) \\
  \tau_1 & \sim & \text{Normal}^+(0, 1) \\
  \tau_2 & \sim & \text{Normal}^+(0, 1) \\
  \sigma & \sim & \text{Normal}^+(0, 1) \\
  \\
  \text{\textit{hierarchical \ priors}} \\
  k_1^n & \sim & \text{Normal}(k_{1, \text{pop}}, \tau_1) \\
  k_2^n & \sim & \text{Normal}(k_{2, \text{pop}}, \tau_2) \\
  \\
  \text{\textit{likelihood}} \\
  y_n & \sim & \text{Normal} \left (m_\text{cent} (t, k_1^n, k_2^n), \sigma \right),
\end{eqnarray*}
where the second argument for the Normal distribution is the standard deviation
and Normal$^+$ is a normal distribution truncated at 0, 
with non-zero density only over positive values.
$m_\text{cent}$ is computed by solving the ordinary differential equation (ODE),
\begin{eqnarray*}
  \frac{\mathrm d m_\text{gut}}{\mathrm d t} & = & - k_1 m_\text{gut} \nonumber  \\
  \frac{\mathrm d m_\text{cent}}{\mathrm d t} & = & k_1 m_\text{gut} - k_2 m_\text{cent},
\end{eqnarray*}
which admits an analytical solution, when $k_1 \neq k_2$,
\begin{eqnarray*}
  m_\text{gut} (t) & = & m^0_\text{gut} \exp(- k_1 t) \nonumber \\
  m_\text{cent} (t) & = & \frac{\exp(- k_2 t)}{k_1 - k_2} \left (m^0_\text{gut} k_1 (1 - \exp[(k_2 - k_1) t] + (k_1 - k_2) m^0_\text{cent}) \right).
\end{eqnarray*}
Here $m^0_\text{gut}$ and $m^0_\text{cent}$ are the initial conditions at time $t = 0$.
Each patient receives one dose at time $t = 0$, 
and measurements are taken at times $t = (0.083, 0.167, 0.25, 1, 2, 4)$ for a total of 6 observations per patient.
Data is simulated over 10 patients. 

In our notation for LGMs,
\begin{eqnarray*}
  \phi = & (\tau_1, \tau_2), & \hspace{1in} \text{\textit{Hyperparameters for prior covariance}} \\
  \eta = & \sigma, & \hspace{1in} \text{\textit{Hyperparameters for likelihood}} \\
  \theta = & ({\bf k}_1, {\bf k}_2), & \hspace{1in} \text{\textit{Latent Gaussian variable}}
\end{eqnarray*}
where ${\bf k}_i$ is a vector with the patient level absorption and clearance parameters.
This model violates all three assumptions for the classical integrated Laplace approximation (Section~\ref{sec:classical-limitation}).
Indeed, $\eta \neq \emptyset$, the Hessian is block-diagonal, with block size $2 \times 2$ (once we organize the parameters to minimize the block size),
and the likelihood is not log-concave.
$W$ does not admit a matrix square-root and the $B$-matrix, \mbox{$B = I + W^\frac{1}{2} K W^\frac{1}{2}$}, cannot be used.
The alternative $B$-matrices both work, with $B =  I + KW$ being slightly more stable.
In both cases, I use a linesearch step.

The optimization problem underlying the Laplace approximation is difficult,
resulting in slow computation and numerical instability as the Markov chain explores the parameter space.
This problem is particularly acute during the warmup phase.
%
%
By contrast, the optimizer behaves reasonably well in the examples studied by \citet{Margossian:2020-laplace}, all of which employ a general linear model likelihood.

\begin{figure}
  \includegraphics[width=6in]{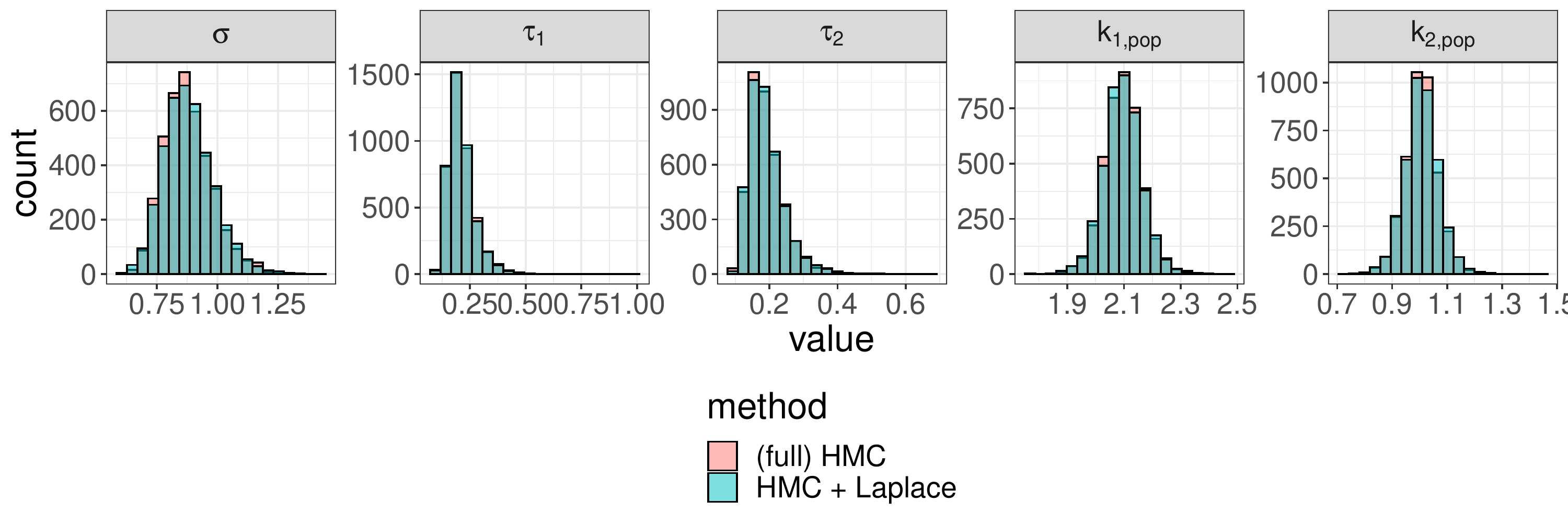}
  \caption{Posterior samples obtained with full HMC and the integrated Laplace approximation on a population pharmacokinetic model}
  \label{fig:PK_sample_comp}
\end{figure}

\begin{figure}
  \begin{center}
  \includegraphics[width=6in]{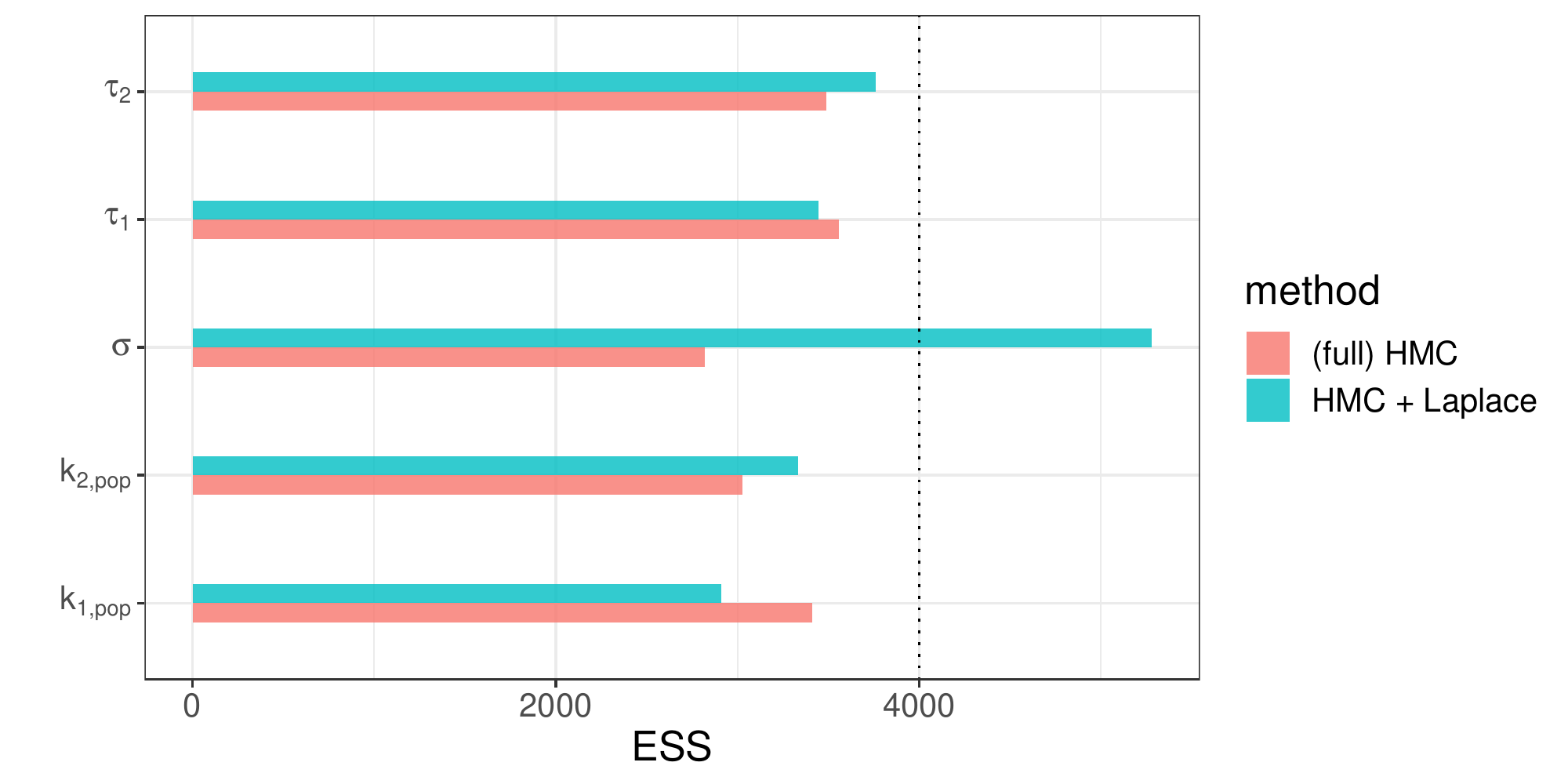}
  \caption{Effective sample size obtained with full HMC and the integrated Laplace approximation on a population pharmacokinetic model. \textit{The dotted line represents the actual sample size.}}
  \label{fig:PK_ESS_comp}
  \end{center}
\end{figure}

\begin{figure}
  \begin{center}
  \includegraphics[width=6in]{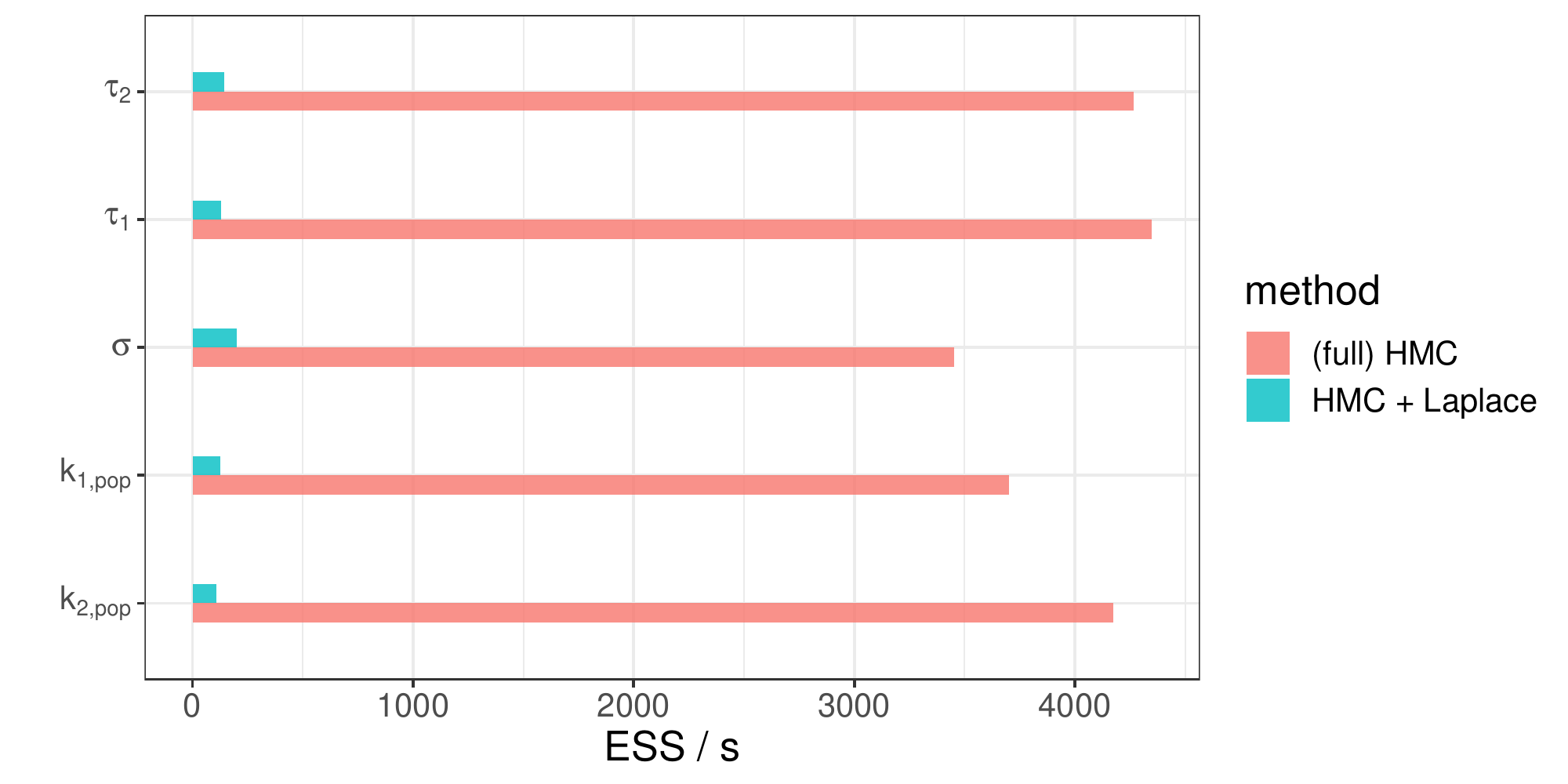}
  \caption{Effective sample size per second obtained with full HMC and the integrated Laplace approximation on a population pharmacokinetic model}
  \label{fig:PK_EFF_comp}
  \end{center}
\end{figure}

I use HMC applied to the full parameter space, or \textit{full HMC}, as a benchmark.
Full HMC does quite well on this example, meaning that, unlike in other cases, the posterior geometry is well-behaved.
Despite the unorthodox nature of the likelihood, the integrated Laplace approximation produces posterior estimates of the hyperparameters which are in close agreement with full HMC (Figure~\ref{fig:PK_sample_comp}).
The bias introduced by the approximation is negligible,
meaning we can compare the effective sample size (ESS) estimated for both samplers.
With both methods, the chain's autocorrelation is relatively small.
The integrated Laplace approximation generates a larger ESS, exceeding the actual sample size for $\sigma$, meaning the Monte Carlo estimators are super-efficient (Figure~\ref{fig:PK_ESS_comp}).
This is because marginalization allows us to run HMC on a posterior distribution with a well-behaved geometry.
That said, the excessively slow optimization for this non-convex problem means the performance of full HMC is vastly superior as measured by the ESS / second (Figure~\ref{fig:PK_EFF_comp}).
More generally, in cases where the posterior geometry does not frustrate MCMC,
we may expect full HMC to outperform the integrated Laplace approximation,
especially if the underlying optimization problem is difficult.

\section{Discussion}

I propose a generalization of the adjoint-differentiated Laplace approximation by (i) expanding the algorithm to work on three Newton solvers, using $B$-matrices as a unifying framework, and (ii) fully automating the differentiation of the likelihood.
The resulting implementation is more flexible and slightly faster than the original method
which uses analytical derivatives.
This greatly facilitates implementing the method in software for a broad range of likelihoods.
The proposed implementation also makes it straightforward to explore less conventional models
such as the population pharmacokinetic model in Section~\ref{sec:gam-experiment}.

Once we consider a rich enough space of models, it becomes clear that the integrated Laplace approximation confronts us to three challenges:
\begin{itemize}
  \item[] (i) \textit{Quality of the approximation}. As we move away from log-concave likelihoods and the theory that supports them, how can we asses whether the approximation is reasonable?
  The population pharmacokinetic example showcases the approximation can be good in an unorthodox setting.
  In other cases with a non-log-concave likelihood, the conditional posterior of the latent variable can be multimodal and therefore not well approximated by a Laplace approximation;
  this problem typically also incurs challenges when computing the approximation
  \citep[e.g.][]{Vanhatalo:2009}.
  Developing inexpensive diagnostic tools to confirm this without running a golden benchmark remains an open problem.
  Candidate diagnostics include importance sampling \citep{Vehtari+etal:2019:psis}, 
  leave-one-out cross-validation \citep{Vehtari+etal:2016:loo_glvm}, and simulation based calibration \citep{Talts:2018}.
  \item[]
  \item[] (ii) \textit{Optimization}. Can we efficiently compute the Laplace approximation? The answer to this question varies between likelihoods, and furthermore the hyperparameter values we encounter as we run MCMC.
  \item[]
  \item[] (iii) \textit{Differentiation}. One persistent limitation is that any operation used to evaluate the log likelihood must support both forward and reverse mode automatic differentiation in order to compute higher-order derivatives.
For implicit functions, we will likely need higher-order adjoint methods to insure an efficient implementation.
\end{itemize}

The general adjoint-differentiated Laplace approximation is prototyped in an experimental branch of \texttt{Stan}.
It is straightforward to embed the Laplace approximation in HMC, as I did in Section~\ref{sec:gam-experiment}, and furthermore in any gradient-based inference algorithm supported by \texttt{Stan}
including penalized optimization, automatic differentiation variational inference \citep{Kucukelbir:2017}, and the prototype pathfinder \citep{Zhang:2021}.
Studying the use of different inference algorithms for the hyperparameter presents an exciting avenue of research.

\section{Code}

The prototype general adjoint-differentiated Laplace approximation for the \texttt{Stan-math} \texttt{C++} library can be found at \url{https://github.com/Stan-dev/math}, under the branch \texttt{experimental/laplace}.
Code to expose the suite of Laplace functions to the \texttt{Stan} language can be found at
\url{https://github.com/Stan-dev/Stanc3/tree/update/laplace-rng}.
Instructions on installing \texttt{Stan} with the relevant branches, along with several examples, can be found at \url{https://github.com/SteveBronder/laplace_testing}.

\section*{Acknowledgment}

I am grateful to Aki Vehtari, Jarno Vanhatalo, and Dan Simpson for many helpful discussions.
I am indebted to Steve Bronder who reviewed my \texttt{C++} code
and refactored it in order to improve the API and its overall quality.
I thank Ben Bales for helping me understand the use of higher-order automatic differentiation in \texttt{Stan},
and Rok \u{C}e\u{s}novar for his help with \texttt{Stan}'s transpiler.


\bibliography{references}

\begin{thebibliography}{35}

\bibitem[\protect\citeauthoryear{Baydin et~al.}{2018}]{Baydin:2018}
\begin{barticle}[author]
\bauthor{\bsnm{Baydin},~\bfnm{Atilim~Gunes}\binits{A.~G.}},
  \bauthor{\bsnm{Pearlmutter},~\bfnm{Barak~A.}\binits{B.~A.}},
  \bauthor{\bsnm{Radul},~\bfnm{Alexey~Andreyevich}\binits{A.~A.}} \AND
  \bauthor{\bsnm{Siskind},~\bfnm{Jeffrey~Mark}\binits{J.~M.}}
(\byear{2018}).
\btitle{Automatic differentiation in machine learning: a survey}.
\bjournal{Journal of Machine Learning Research}
\bvolume{18}
\bpages{1 -- 43}.
\end{barticle}
\endbibitem

\bibitem[\protect\citeauthoryear{Betancourt and
  Girolami}{2015}]{Betancourt:2015}
\begin{binbook}[author]
\bauthor{\bsnm{Betancourt},~\bfnm{Michael}\binits{M.}} \AND
  \bauthor{\bsnm{Girolami},~\bfnm{Mark}\binits{M.}}
(\byear{2015}).
\btitle{Current Trends in Bayesian Methodology with Applications}
\bchapter{{Hamiltonian Monte Carlo} for Hierarchical Models}.
\bpublisher{Chapman and Hall/CRC}.
\bdoi{10.1201/b18502-5}
\end{binbook}
\endbibitem

\bibitem[\protect\citeauthoryear{Carpenter et~al.}{2015}]{Carpenter:2015}
\begin{barticle}[author]
\bauthor{\bsnm{Carpenter},~\bfnm{Bob}\binits{B.}},
  \bauthor{\bsnm{Hoffman},~\bfnm{Matthew~D.}\binits{M.~D.}},
  \bauthor{\bsnm{Brubaker},~\bfnm{Marcus~A.}\binits{M.~A.}},
  \bauthor{\bsnm{Lee},~\bfnm{Daniel}\binits{D.}},
  \bauthor{\bsnm{Li},~\bfnm{Peter}\binits{P.}} \AND
  \bauthor{\bsnm{Betancourt},~\bfnm{Michael~J.}\binits{M.~J.}}
(\byear{2015}).
\btitle{The {S}tan Math Library: Reverse-Mode Automatic Differentiation in
  {C}++}.
\bjournal{arXiv 1509.07164.}
\end{barticle}
\endbibitem

\bibitem[\protect\citeauthoryear{Carpenter et~al.}{2017}]{Carpenter:2017}
\begin{barticle}[author]
\bauthor{\bsnm{Carpenter},~\bfnm{Bob}\binits{B.}},
  \bauthor{\bsnm{Gelman},~\bfnm{Andrew}\binits{A.}},
  \bauthor{\bsnm{Hoffman},~\bfnm{Matt}\binits{M.}},
  \bauthor{\bsnm{Lee},~\bfnm{Daniel}\binits{D.}},
  \bauthor{\bsnm{Goodrich},~\bfnm{Ben}\binits{B.}},
  \bauthor{\bsnm{Betancourt},~\bfnm{Michael}\binits{M.}},
  \bauthor{\bsnm{Brubaker},~\bfnm{Marcus~A.}\binits{M.~A.}},
  \bauthor{\bsnm{Guo},~\bfnm{Jiqiang}\binits{J.}},
  \bauthor{\bsnm{Li},~\bfnm{Peter}\binits{P.}} \AND
  \bauthor{\bsnm{Riddel},~\bfnm{Allen}\binits{A.}}
(\byear{2017}).
\btitle{Stan: {A} Probabilistic Programming Language}.
\bjournal{Journal of Statistical Software}
\bvolume{76}
\bpages{1 --32}.
\bdoi{10.18637/jss.v076.i01}
\end{barticle}
\endbibitem

\bibitem[\protect\citeauthoryear{Gaebler}{2021}]{Gaebler:2021}
\begin{bmisc}[author]
\bauthor{\bsnm{Gaebler},~\bfnm{Johann~D}\binits{J.~D.}}
(\byear{2021}).
\btitle{Autodiff for Implicit Functions in Stan}.
\end{bmisc}
\endbibitem

\bibitem[\protect\citeauthoryear{Gastonguay and {Metrum Institute
  Facility}}{2013}]{Gastonguay:2013}
\begin{bbook}[author]
\bauthor{\bsnm{Gastonguay},~\bfnm{Marc}\binits{M.}} \AND \bauthor{\bsnm{{Metrum
  Institute Facility}}}
(\byear{2013}).
\btitle{{MI-210}: Essentials of Population {PKPD} Modeling and Simulation}.
\end{bbook}
\endbibitem

\bibitem[\protect\citeauthoryear{Gibaldi and Perrier}{1982}]{Gibaldi:1982}
\begin{bbook}[author]
\bauthor{\bsnm{Gibaldi},~\bfnm{Milo}\binits{M.}} \AND
  \bauthor{\bsnm{Perrier},~\bfnm{Donald}\binits{D.}}
(\byear{1982}).
\btitle{Pharmacokinetics},
\bedition{2} ed.
\end{bbook}
\endbibitem

\bibitem[\protect\citeauthoryear{G\'omez-Rubio and Rue}{2018}]{Gomez:2018}
\begin{barticle}[author]
\bauthor{\bsnm{G\'omez-Rubio},~\bfnm{Virgilio}\binits{V.}} \AND
  \bauthor{\bsnm{Rue},~\bfnm{Havard}\binits{H.}}
(\byear{2018}).
\btitle{Markov chain {Monte Carlo} with the Integrated Nested {Laplace}
  Approximation}.
\bjournal{Statistics and Computing}
\bvolume{28}
\bpages{1033 -- 1051}.
\end{barticle}
\endbibitem

\bibitem[\protect\citeauthoryear{Gower and Mello}{2011}]{Gower:2011}
\begin{barticle}[author]
\bauthor{\bsnm{Gower},~\bfnm{R.~M.}\binits{R.~M.}} \AND
  \bauthor{\bsnm{Mello},~\bfnm{M.~P.}\binits{M.~P.}}
(\byear{2011}).
\btitle{A new framework for the computation of {H}essians}.
\bjournal{Optimization methods and software}
\bvolume{27}
\bpages{251--273}.
\bdoi{https://doi.org/10.1080/10556788.2011.580098}
\end{barticle}
\endbibitem

\bibitem[\protect\citeauthoryear{Griewank and Walther}{2008}]{Griewank:2008}
\begin{bbook}[author]
\bauthor{\bsnm{Griewank},~\bfnm{Andreas}\binits{A.}} \AND
  \bauthor{\bsnm{Walther},~\bfnm{Andrea}\binits{A.}}
(\byear{2008}).
\btitle{Evaluating derivatives},
\bedition{Second} ed.
\bpublisher{Society for Industrial and Applied Mathematics ({SIAM}),
  Philadelphia, PA}.
\end{bbook}
\endbibitem

\bibitem[\protect\citeauthoryear{Joensuu et~al.}{2012}]{Joensuu:2012}
\begin{barticle}[author]
\bauthor{\bsnm{Joensuu},~\bfnm{Heikki}\binits{H.}},
  \bauthor{\bsnm{Vehtari},~\bfnm{Aki}\binits{A.}},
  \bauthor{\bsnm{Riihimäki},~\bfnm{Jaakko}\binits{J.}},
  \bauthor{\bsnm{Nishida},~\bfnm{Toshirou}\binits{T.}},
  \bauthor{\bsnm{Steigen},~\bfnm{Sonja~E}\binits{S.~E.}},
  \bauthor{\bsnm{Brabec},~\bfnm{Peter}\binits{P.}},
  \bauthor{\bsnm{Plank},~\bfnm{Lukas}\binits{L.}},
  \bauthor{\bsnm{Nilsson},~\bfnm{Bengt}\binits{B.}},
  \bauthor{\bsnm{Cirilli},~\bfnm{Claudia}\binits{C.}},
  \bauthor{\bsnm{Braconi},~\bfnm{Chiara}\binits{C.}},
  \bauthor{\bsnm{Bordoni},~\bfnm{Andrea}\binits{A.}},
  \bauthor{\bsnm{Magnusson},~\bfnm{Magnus~K}\binits{M.~K.}},
  \bauthor{\bsnm{Linke},~\bfnm{Zdenek}\binits{Z.}},
  \bauthor{\bsnm{Sufliarsky},~\bfnm{Jozef}\binits{J.}},
  \bauthor{\bsnm{Federico},~\bfnm{Massimo}\binits{M.}},
  \bauthor{\bsnm{Jonasson},~\bfnm{Jon~G}\binits{J.~G.}}, \bauthor{\bsnm{{Dei
  Tos}},~\bfnm{Angelo~Paolo}\binits{A.~P.}} \AND
  \bauthor{\bsnm{Rutkowski},~\bfnm{Piotr}\binits{P.}}
(\byear{2012}).
\btitle{Risk of recurrence of gastrointestinal stromal tumour after surgery: an
  analysis of pooled population-based cohorts}.
\bjournal{The Lancet Oncology}
\bvolume{13}
\bpages{265-274}.
\bdoi{https://doi.org/10.1016/S1470-2045(11)70299-6}
\end{barticle}
\endbibitem

\bibitem[\protect\citeauthoryear{Jyl\"anki, Vanhatalo and
  Vehtari}{2011}]{Jylanki:2011}
\begin{barticle}[author]
\bauthor{\bsnm{Jyl\"anki},~\bfnm{Pasi}\binits{P.}},
  \bauthor{\bsnm{Vanhatalo},~\bfnm{Jarno}\binits{J.}} \AND
  \bauthor{\bsnm{Vehtari},~\bfnm{Aki}\binits{A.}}
(\byear{2011}).
\btitle{Robust {G}aussian Process Regression with a {S}tudent-$t$ Likelihood}.
\bjournal{Journal of Machine Learning Research}
\bvolume{12}
\bpages{3227 -- 3257}.
\end{barticle}
\endbibitem

\bibitem[\protect\citeauthoryear{Kristensen et~al.}{2016}]{Kristensen:2016}
\begin{barticle}[author]
\bauthor{\bsnm{Kristensen},~\bfnm{Kasper}\binits{K.}},
  \bauthor{\bsnm{Nielsen},~\bfnm{Anders}\binits{A.}},
  \bauthor{\bsnm{Berg},~\bfnm{Casper~W}\binits{C.~W.}},
  \bauthor{\bsnm{Skaug},~\bfnm{Hans}\binits{H.}} \AND
  \bauthor{\bsnm{Bell},~\bfnm{Bradley~M}\binits{B.~M.}}
(\byear{2016}).
\btitle{{TMB}: Automatic Differentiation and {L}aplace Approximation}.
\bjournal{Journal of statistical software}
\bvolume{70}
\bpages{1 -- 21}.
\end{barticle}
\endbibitem

\bibitem[\protect\citeauthoryear{Kucukelbir et~al.}{2017}]{Kucukelbir:2017}
\begin{barticle}[author]
\bauthor{\bsnm{Kucukelbir},~\bfnm{Alp}\binits{A.}},
  \bauthor{\bsnm{Tran},~\bfnm{Dustin}\binits{D.}},
  \bauthor{\bsnm{Ranganath},~\bfnm{Rajesh}\binits{R.}},
  \bauthor{\bsnm{Gelman},~\bfnm{Andrew}\binits{A.}} \AND
  \bauthor{\bsnm{Blei},~\bfnm{David}\binits{D.}}
(\byear{2017}).
\btitle{Automatic differentiation variational inference}.
\bjournal{Journal of machine learning research}
\bvolume{18}
\bpages{1 -- 45}.
\end{barticle}
\endbibitem

\bibitem[\protect\citeauthoryear{Margossian}{2019}]{Margossian:2019}
\begin{barticle}[author]
\bauthor{\bsnm{Margossian},~\bfnm{Charles~C.}\binits{C.~C.}}
(\byear{2019}).
\btitle{A Review of automatic differentiation and its efficient
  implementation}.
\bjournal{Wiley interdisciplinary reviews: data mining and knowledge discovery}
\bvolume{9}.
\bdoi{10.1002/WIDM.1305}
\end{barticle}
\endbibitem

\bibitem[\protect\citeauthoryear{Margossian}{2022}]{Margossian:2022}
\begin{bbook}[author]
\bauthor{\bsnm{Margossian},~\bfnm{Charles~C}\binits{C.~C.}}
(\byear{2022}).
\btitle{Modernizing {Markov chains Monte Carlo} for scientific and {Bayesian}
  modeling}.
\bnote{PhD Thesis}.
\end{bbook}
\endbibitem

\bibitem[\protect\citeauthoryear{Margossian and
  Betancourt}{2022}]{Margossian:2022-autodiff}
\begin{barticle}[author]
\bauthor{\bsnm{Margossian},~\bfnm{Charles~C}\binits{C.~C.}} \AND
  \bauthor{\bsnm{Betancourt},~\bfnm{Michael}\binits{M.}}
(\byear{2022}).
\btitle{Efficient Automatic Differentiation of Implicit Functions}.
\bjournal{arXiv:2112.14217}.
\end{barticle}
\endbibitem

\bibitem[\protect\citeauthoryear{Margossian, Zhang and
  Gillespie}{2022}]{Margossian:2021-torsten}
\begin{barticle}[author]
\bauthor{\bsnm{Margossian},~\bfnm{Charles~C}\binits{C.~C.}},
  \bauthor{\bsnm{Zhang},~\bfnm{Yi}\binits{Y.}} \AND
  \bauthor{\bsnm{Gillespie},~\bfnm{William~R}\binits{W.~R.}}
(\byear{2022}).
\btitle{Flexible and efficient Bayesian pharmacometrics modeling using Stan and
  Torsten, Part I}.
\bjournal{CPT: Pharmacometrics \& Systems Pharmacology}
\bvolume{11}
\bpages{1151 -- 1169}.
\bdoi{https://doi.org/10.1002/psp4.12812}
\end{barticle}
\endbibitem

\bibitem[\protect\citeauthoryear{Margossian
  et~al.}{2020}]{Margossian:2020-laplace}
\begin{barticle}[author]
\bauthor{\bsnm{Margossian},~\bfnm{Charles~C}\binits{C.~C.}},
  \bauthor{\bsnm{Vehtari},~\bfnm{Aki}\binits{A.}},
  \bauthor{\bsnm{Simpson},~\bfnm{Daniel}\binits{D.}} \AND
  \bauthor{\bsnm{Agrawal},~\bfnm{Raj}\binits{R.}}
(\byear{2020}).
\btitle{{Hamiltonian Monte Carlo} using an adjoint-differentiated {L}aplace
  approximation: {B}ayesian inference for latent Gaussian models and beyond}.
\bjournal{Neural Information Processing Systems}.
\end{barticle}
\endbibitem

\bibitem[\protect\citeauthoryear{Monnahan and Kristensen}{2018}]{Monnahan:2018}
\begin{barticle}[author]
\bauthor{\bsnm{Monnahan},~\bfnm{Cole~C}\binits{C.~C.}} \AND
  \bauthor{\bsnm{Kristensen},~\bfnm{Kasper}\binits{K.}}
(\byear{2018}).
\btitle{{No-U-turn} sampling for fast {Bayesian} inference in {ADMB} and {TMB}:
  Introducing the adnuts and tmbstan {R} packages}.
\bjournal{Plos One}
\bvolume{13}.
\bdoi{https://doi.org/10.1371/journal.pone.0197954}
\end{barticle}
\endbibitem

\bibitem[\protect\citeauthoryear{Rasmussen and Nickish}{2010}]{Rasmussen:2010}
\begin{barticle}[author]
\bauthor{\bsnm{Rasmussen},~\bfnm{C.~E.}\binits{C.~E.}} \AND
  \bauthor{\bsnm{Nickish},~\bfnm{Hannes}\binits{H.}}
(\byear{2010}).
\btitle{Gaussian processes for machine learning ({GPML}) toolbox}.
\bjournal{Journal of Machine Learning Research}
\bvolume{11}
\bpages{3011 -- 3015}.
\end{barticle}
\endbibitem

\bibitem[\protect\citeauthoryear{Rasmussen and Williams}{2006}]{Rasmussen:2006}
\begin{bbook}[author]
\bauthor{\bsnm{Rasmussen},~\bfnm{C.~E.}\binits{C.~E.}} \AND
  \bauthor{\bsnm{Williams},~\bfnm{C.~K.~I.}\binits{C.~K.~I.}}
(\byear{2006}).
\btitle{Gaussian Processes for Machine Learning}.
\bpublisher{The MIT Press}.
\end{bbook}
\endbibitem

\bibitem[\protect\citeauthoryear{Riihim\"aki and
  Vehtari}{2014}]{Riihimaki:2014}
\begin{barticle}[author]
\bauthor{\bsnm{Riihim\"aki},~\bfnm{Jaakko}\binits{J.}} \AND
  \bauthor{\bsnm{Vehtari},~\bfnm{Aki}\binits{A.}}
(\byear{2014}).
\btitle{Laplace Approximation for Logistic {G}aussian Process Density
  Estimation and Regression}.
\bjournal{Bayesian analysis}
\bvolume{9}
\bpages{425 -- 448}.
\bdoi{10.1214/14-BA872}
\end{barticle}
\endbibitem

\bibitem[\protect\citeauthoryear{Rue, Martino and Chopin}{2009}]{Rue:2009}
\begin{barticle}[author]
\bauthor{\bsnm{Rue},~\bfnm{Havard}\binits{H.}},
  \bauthor{\bsnm{Martino},~\bfnm{Sara}\binits{S.}} \AND
  \bauthor{\bsnm{Chopin},~\bfnm{Nicolas}\binits{N.}}
(\byear{2009}).
\btitle{Approximate {B}ayesian inference for latent {G}aussian models by using
  integrated nested {L}aplace approximations}.
\bjournal{Journal of Royal Statistics B}
\bvolume{71}
\bpages{319 -- 392}.
\end{barticle}
\endbibitem

\bibitem[\protect\citeauthoryear{Rue et~al.}{2017}]{Rue:2017}
\begin{barticle}[author]
\bauthor{\bsnm{Rue},~\bfnm{Havard}\binits{H.}},
  \bauthor{\bsnm{Riebler},~\bfnm{Andrea}\binits{A.}},
  \bauthor{\bsnm{Sorbye},~\bfnm{Sigrunn}\binits{S.}},
  \bauthor{\bsnm{Illian},~\bfnm{Janine}\binits{J.}},
  \bauthor{\bsnm{Simson},~\bfnm{Daniel}\binits{D.}} \AND
  \bauthor{\bsnm{Lindgren},~\bfnm{Finn}\binits{F.}}
(\byear{2017}).
\btitle{Bayesian Computing with {INLA}: A Review}.
\bjournal{Annual Review of Statistics and its Application}
\bvolume{4}
\bpages{395 -- 421}.
\bdoi{https://doi.org/10.1146/annurev-statistics-060116-054045}
\end{barticle}
\endbibitem

\bibitem[\protect\citeauthoryear{Talts et~al.}{2020}]{Talts:2018}
\begin{barticle}[author]
\bauthor{\bsnm{Talts},~\bfnm{Sean}\binits{S.}},
  \bauthor{\bsnm{Betancourt},~\bfnm{Michael}\binits{M.}},
  \bauthor{\bsnm{Simpson},~\bfnm{Daniel}\binits{D.}},
  \bauthor{\bsnm{Vehtari},~\bfnm{Aki}\binits{A.}} \AND
  \bauthor{\bsnm{Gelman},~\bfnm{Andrew}\binits{A.}}
(\byear{2020}).
\btitle{Validating {Bayesian} inference algorithms with simulation-based
  calibration}.
\bjournal{arXiv:1804.06788v1}.
\end{barticle}
\endbibitem

\bibitem[\protect\citeauthoryear{Tierney and Kadane}{1986}]{Tierney:1986}
\begin{barticle}[author]
\bauthor{\bsnm{Tierney},~\bfnm{Luke}\binits{L.}} \AND
  \bauthor{\bsnm{Kadane},~\bfnm{Joseph~B.}\binits{J.~B.}}
(\byear{1986}).
\btitle{Accurate Approximations for Posterior Moments and Marginal Densities}.
\bjournal{Journal of the American Statistical Association}
\bvolume{81}
\bpages{82-86}.
\bdoi{10.1080/01621459.1986.10478240}
\end{barticle}
\endbibitem

\bibitem[\protect\citeauthoryear{Tolvanen, Jyl\"anki and
  Vehtari}{2014}]{Tolvanen:2014}
\begin{barticle}[author]
\bauthor{\bsnm{Tolvanen},~\bfnm{Ville}\binits{V.}},
  \bauthor{\bsnm{Jyl\"anki},~\bfnm{Pasi}\binits{P.}} \AND
  \bauthor{\bsnm{Vehtari},~\bfnm{Aki}\binits{A.}}
(\byear{2014}).
\btitle{Expectation propagation for nonstationary heteroscedastic {G}aussian
  process regression}.
\bjournal{Machine Learning for Signal Processing (MLSP), 2014 IEEE
  International Workshop on}.
\bdoi{doi:10.1109/MLSP.2014.6958906.}
\end{barticle}
\endbibitem

\bibitem[\protect\citeauthoryear{Vanhatalo, Foster and
  Hosack}{2021}]{Vanhatalo:2021}
\begin{barticle}[author]
\bauthor{\bsnm{Vanhatalo},~\bfnm{Jarno}\binits{J.}},
  \bauthor{\bsnm{Foster},~\bfnm{Scott~D.}\binits{S.~D.}} \AND
  \bauthor{\bsnm{Hosack},~\bfnm{Geoffrey}\binits{G.}}
(\byear{2021}).
\btitle{Spatiotemporal clustering using {G}aussian processes embedded in a
  mixture model}.
\bjournal{Environmetrics}
\bvolume{32}.
\end{barticle}
\endbibitem

\bibitem[\protect\citeauthoryear{Vanhatalo, Jyl\"{a}nki and
  Vehtari}{2009}]{Vanhatalo:2009}
\begin{barticle}[author]
\bauthor{\bsnm{Vanhatalo},~\bfnm{Jarno}\binits{J.}},
  \bauthor{\bsnm{Jyl\"{a}nki},~\bfnm{Pasi}\binits{P.}} \AND
  \bauthor{\bsnm{Vehtari},~\bfnm{Aki}\binits{A.}}
(\byear{2009}).
\btitle{Gaussian process regression with {S}tudent-$t$ likelihood}.
\end{barticle}
\endbibitem

\bibitem[\protect\citeauthoryear{Vanhatalo et~al.}{2013}]{Vanhatalo:2013}
\begin{barticle}[author]
\bauthor{\bsnm{Vanhatalo},~\bfnm{Jarno}\binits{J.}},
  \bauthor{\bsnm{Riihim\"aki},~\bfnm{Jaakko}\binits{J.}},
  \bauthor{\bsnm{Hartikainen},~\bfnm{Jouni}\binits{J.}},
  \bauthor{\bsnm{Jyl\"anki},~\bfnm{Pasi}\binits{P.}},
  \bauthor{\bsnm{Tolvanen},~\bfnm{Ville}\binits{V.}} \AND
  \bauthor{\bsnm{Vehtari},~\bfnm{Aki}\binits{A.}}
(\byear{2013}).
\btitle{{GPstuff}: {Bayesian} Modeling with {Gaussian} Processes}.
\bjournal{Journal of Machine Learning Research}
\bvolume{14}
\bpages{1175--1179}.
\end{barticle}
\endbibitem

\bibitem[\protect\citeauthoryear{Vehtari}{2021}]{Vehtari:2021}
\begin{bmisc}[author]
\bauthor{\bsnm{Vehtari},~\bfnm{Aki}\binits{A.}}
(\byear{2021}).
\btitle{Gaussian process demonstration with {S}tan}.
\end{bmisc}
\endbibitem

\bibitem[\protect\citeauthoryear{Vehtari
  et~al.}{2016}]{Vehtari+etal:2016:loo_glvm}
\begin{barticle}[author]
\bauthor{\bsnm{Vehtari},~\bfnm{Aki}\binits{A.}},
  \bauthor{\bsnm{Mononen},~\bfnm{Tommi}\binits{T.}},
  \bauthor{\bsnm{Tolvanen},~\bfnm{Ville}\binits{V.}},
  \bauthor{\bsnm{Sivula},~\bfnm{Tuomas}\binits{T.}} \AND
  \bauthor{\bsnm{Winther},~\bfnm{Ole}\binits{O.}}
(\byear{2016}).
\btitle{Bayesian Leave-One-Out Cross-Validation Approximations for {Gaussian}
  Latent Variable Models}.
\bjournal{Journal of Machine Learning Research}
\bvolume{17}
\bpages{1--38}.
\end{barticle}
\endbibitem

\bibitem[\protect\citeauthoryear{Vehtari et~al.}{2019}]{Vehtari+etal:2019:psis}
\begin{barticle}[author]
\bauthor{\bsnm{Vehtari},~\bfnm{Aki}\binits{A.}},
  \bauthor{\bsnm{Simpson},~\bfnm{Daniel}\binits{D.}},
  \bauthor{\bsnm{Gelman},~\bfnm{Andrew}\binits{A.}},
  \bauthor{\bsnm{Yao},~\bfnm{Yuling}\binits{Y.}} \AND
  \bauthor{\bsnm{Gabry},~\bfnm{Jonah}\binits{J.}}
(\byear{2019}).
\btitle{Pareto smoothed importance sampling}.
\bjournal{arXiv:1507.02646}.
\end{barticle}
\endbibitem

\bibitem[\protect\citeauthoryear{Zhang et~al.}{2022}]{Zhang:2021}
\begin{barticle}[author]
\bauthor{\bsnm{Zhang},~\bfnm{Lu}\binits{L.}},
  \bauthor{\bsnm{Carpenter},~\bfnm{Bob}\binits{B.}},
  \bauthor{\bsnm{Gelman},~\bfnm{Andrew}\binits{A.}} \AND
  \bauthor{\bsnm{Vehtari},~\bfnm{Aki}\binits{A.}}
(\byear{2022}).
\btitle{Pathfinder: Parallel quasi-{N}ewton variational inference}.
\bjournal{Journal of Machine Learning Research}
\bvolume{23}
\bpages{1 - 49}.
\end{barticle}
\endbibitem

\end{thebibliography}
\bibliographystyle{imsart-nameyear}

\appendix

\section{}

This appendix provides a proof of Proposition~\ref{lemma:diff-eta}, which provides the derivative of the approximate marginal likelihood, $\log \pi_\mathcal{G} \pi(y \mid \phi, \eta)$ with respect to $\eta$.

Following the steps from \citet{Rasmussen:2006} but differentiating with respect to $\eta$,
\begin{equation}
  \frac{\mathrm d }{\mathrm d \eta_j} \log \pi_\mathcal{G}(y \mid \phi) = \frac{\partial \log \pi_\mathcal{G}(y \mid \phi)}{\partial \eta_j} + \sum_i \frac{\partial \log \pi_\mathcal{G}(y \mid \phi)}{\partial \hat \theta_i} \frac{\mathrm d \hat \theta_i}{\mathrm d \eta_j}.
\end{equation}
The first ``explicit'' term is
\begin{equation}
  \frac{\partial \log \pi_\mathcal{G}(y \mid \phi)}{\partial \eta_j}
    = \frac{\partial \log \pi(y \mid \hat \theta, \eta)}{\partial \eta_j} - \frac{1}{2} \frac{\partial \log |K||K + W^{-1}|}{\partial \eta_j}.
\end{equation}
We can work out the first term analytically (or with automatic differentiation).
Next we consider the following handy lemma.
\begin{lemma} \cite[equation A.15]{Rasmussen:2006}
  For an invertible and differentiable natrix $A$, 
\begin{equation*}
  \frac{\partial}{\partial \theta} \log |A| = \mathrm{tr} \left (A^{-1} \frac{\partial A}{\partial \theta} \right).
\end{equation*}
\end{lemma}

Hence
\begin{eqnarray*}
  \frac{\partial \log |K||K^{-1} + W|}{\partial \eta_j} & = & \frac{\partial \log |K^{-1} + W|}{\partial \eta_j} \\
    & = & \mathrm{tr} \left ((K^{-1} + W)^{-1} \frac{\partial W}{\partial \eta_j} \right) \\
    & = & - \mathrm{tr} \left ((K^{-1} + W)^{-1} \frac{\partial \nabla_\theta^2 \log \pi(y \mid \theta, \eta)}{\partial \eta_j} \right)
\end{eqnarray*} 
where we recall that $\eta$ parameterizes $W$ but not $K$.

Now for the implicit term, we differentiate the self-consistent equation
\begin{equation*}
  \hat \theta = K \nabla_\theta \log \pi(y \mid \theta, \eta).
\end{equation*}
Thus
\begin{eqnarray*}
  \frac{\partial \hat \theta}{\partial \eta_l} & = & K \left [ \frac{\partial}{\partial \eta} \nabla_\theta \log \pi(y \mid \theta, \eta) + \frac{\partial \nabla_\theta \log \pi(y \mid \theta, \eta)}{\partial \hat \theta} \frac{\partial \hat \theta}{\partial \eta_l} \right ] \\
  & = & K \left [ \frac{\partial}{\partial \eta} \nabla_\theta \log \pi(y \mid \theta, \eta) - W \frac{\partial \hat \theta}{\partial \eta_l} \right ],
\end{eqnarray*}
equivalently,
\begin{eqnarray*}
  \frac{\partial \hat \theta}{\partial \eta_l} = (I + KW)^{-1} K \frac{\partial}{\partial \eta} \nabla_\theta \log \pi(y \mid \theta, \eta).
\end{eqnarray*}
Putting it all together
\begin{equation*} \label{eq:gradient}
\begin{aligned}
  \frac{\partial \log \pi_\mathcal{G} (y \mid \phi)}{\partial \eta_l} =&  \frac{\partial \log \pi(y \mid \hat \theta, \eta)}{\partial \eta_l} \\
  & 
  + \frac{1}{2} \mathrm{tr} \left ((K^{-1} + W)^{-1} \frac{\partial \nabla^2_\theta \log \pi(y \mid \hat \theta, \eta)}{\partial \eta_j} \right) \\
 & + \sum_{i = 1}^n \frac{\partial \log \pi_\mathcal{G} (y \mid \phi)}{\partial \hat \theta_i} (I + KW)^{-1} K \frac{\partial}{\partial \eta_l} \nabla_\theta \log \pi(y \mid \theta, \eta),
\end{aligned}
\end{equation*}
where 
\begin{eqnarray*}
   \frac{\partial \log \pi_\mathcal{G}(y \mid \phi)}{\partial \hat \theta_i} & = & - \frac{1}{2} \frac{\partial}{\partial \hat \theta_i} \log |K^{-1} + W| \\
   & = & - \frac{1}{2} \mathrm{tr} \left ((K^{-1} + W)^{-1} \frac{\partial W}{\partial \theta_i} \right) \\
   & = & \frac{1}{2} \left [ (K^{-1} + W)^{-1} \right ]_{ii} \frac{\partial^3}{\partial \theta^3_i} \log \pi(y \mid \hat \theta, \eta),
\end{eqnarray*}
with the last line holding only in the special case where the Hessian is diagonal \citep[equation 5.23]{Rasmussen:2006}. \qed

\end{document}